%% file: main.tex
\documentclass[preprint]{elsarticle}

\usepackage{lineno,hyperref}
\modulolinenumbers[1]

\usepackage{amsmath}
\usepackage{amssymb}

\usepackage{soul,xcolor}

\usepackage{booktabs}
\usepackage{geometry}
\usepackage{array}

\usepackage[colorinlistoftodos]{todonotes}

\usepackage{float}

\usepackage{ulem}

\journal{Aperture}

\biboptions{authoryear}

\begin{document}

\begin{frontmatter}

\title{Hemodynamic Deconvolution Demystified: Sparsity-Driven Regularization at Work}

\author[bcbl,upv]{Eneko Uru\~nuela\corref{mycorrespondingauthor}}
\cortext[mycorrespondingauthor]{Corresponding authors}
\ead{e.urunuela@bcbl.eu}

\author[epfl,chuv]{Thomas A.W. Bolton}
\author[epfl,unige]{Dimitri Van De Ville}
\author[bcbl]{C\'{e}sar Caballero-Gaudes\corref{mycorrespondingauthor}}
\ead{c.caballero@bcbl.eu}

\address[bcbl]{Basque Center on Cognition, Brain and Language (BCBL),
Donostia-San Sebasti\'{a}n, Spain.} \address[upv]{University of the Basque
Country (EHU/UPV), Donostia-San Sebasti\'{a}n, Spain.}
\address[epfl]{Ecole Polytechnique F\'ed\'erale de Lausanne (EPFL), Lausanne, Switzerland.}
\address[chuv]{Gamma Knife Center, Department of Clinical Neuroscience, Centre Hospitalier Universitaire Vaudois (CHUV), Lausanne, Switzerland}
\address[unige]{Faculty of Medicine, University of Geneva, Geneva, Switzerland}

\begin{abstract}
\input{sections/abstract.tex}
\end{abstract}


\end{frontmatter}


\input{sections/introduction.tex}

\input{sections/theory.tex}

\input{sections/methods.tex}

\input{sections/results.tex}

\input{sections/discussion.tex}

\section{Code and data availability}
\label{sec:github}
The code and materials used in this work can be found in the following GitHub
repository: \url{https://github.com/eurunuela/pfm_vs_ta}. We encourage the
reader to explore the parameters (e.g., SNR, varying HRF options and mismatch
between algorithms, TR, number of events, onsets, and durations) in the provided
Jupyter notebooks. Likewise, the data used to produce the figures can be found
in \url{https://osf.io/f3ryg/}.

\section{Acknowledgements}
We thank Stefano Moia and Vicente Ferrer for data availability, and Younes
Farouj for valuable comments on the manuscript. This research was funded by the
Spanish Ministry of Economy and Competitiveness (RYC-2017-21845), the Basque
Government (BERC 2018-2021, PIB\_2019\_104, PRE\_2020\_2\_0227), and the Spanish
Ministry of Science, Innovation and Universities (PID2019-105520GB-100), and the
Swiss National Science Foundation (grant 205321\_163376).

\section{CRediT}
Eneko Uru\~nuela: Conceptualisation, Methodology, Software, Formal Analysis,
Investigation, Data Curation, Writing (OD), Writing (RE), Visualisation, Funding
acquisition. Thomas A.W. Bolton: Conceptualisation, Methodology, Writing (RE).
Dimitri Van de Ville: Conceptualisation, Methodology, Writing (RE). C\'{e}sar
Caballero-Gaudes: Conceptualisation, Methodology, Software, Formal Analysis,
Investigation, Data Curation, Writing (OD), Writing (RE), Visualisation, Funding
acquisition.

\bibliography{mybibfile}

\newpage
\input{sections/supplementary.tex}

\end{document}

%% file: sections/abstract.tex
Deconvolution of the hemodynamic response is an important step to access short
timescales of brain activity recorded by functional magnetic resonance imaging
(fMRI). Albeit conventional deconvolution algorithms have been around for a long
time (e.g., Wiener deconvolution), recent state-of-the-art methods based on
sparsity-pursuing regularization are attracting increasing interest to
investigate brain dynamics and connectivity with fMRI. This technical note
revisits the main concepts underlying two main methods, Paradigm Free Mapping
and Total Activation, in the most accessible way. Despite their apparent
differences in the formulation, these methods are theoretically equivalent as
they represent the synthesis and analysis sides of the same problem,
respectively. We demonstrate this equivalence in practice with their
best-available implementations using both simulations, with different
signal-to-noise ratios, and experimental fMRI data acquired during a motor task
and resting-state. We evaluate the parameter settings that lead to equivalent
results, and showcase the potential of these algorithms compared to other common
approaches. This note is useful for practitioners interested in gaining a better
understanding of state-of-the-art hemodynamic deconvolution, and aims to answer
questions that practitioners often have regarding the differences between the
two methods.

%% file: sections/introduction.tex

\section{Introduction}

Functional magnetic resonance imaging (fMRI) data analysis is often directed to
identify and disentangle the neural processes that occur in different brain
regions during task or at rest. As the blood oxygenation level-dependent (BOLD)
signal of fMRI is only a proxy for neuronal activity mediated through
neurovascular coupling, an intermediate step that estimates the
activity-inducing signal, at the timescale of fMRI, from the BOLD timeseries can
be useful. Conventional analysis of task fMRI data relies on the general linear
models (GLM) to establish statistical parametric maps of brain activity by
regression of the empirical timecourses against hypothetical ones built from the
knowledge of the experimental paradigm. However, timing information of the
paradigm can be unknown, inaccurate, or insufficient in some scenarios such as
naturalistic stimuli, resting-state, or clinically-relevant assessments.

Deconvolution and methods alike are aiming to estimate neuronal activity by
undoing the blurring effect of the hemodynamic response, characterized as a
hemodynamic response function (HRF)\footnote{Note that the term
deconvolution is also alternatively employed to refer to the estimation of the
hemodynamic response shape assuming a known activity-inducing signal or neuronal
activity
\citep{Goutte2000Modelinghaemodynamicresponse,Marrelec2002Bayesianestimationhemodynamic,
Ciuciu2003Unsupervisedrobustnonparametric,Casanova2008impacttemporalregularization}.
}. Given the inherently ill-posed nature of
hemodynamic deconvolution, due to the strong temporal low-pass characteristics
of the HRF, the key is to introduce additional constraints in the estimation
problem that are typically expressed as regularizers. For instance, the
so-called Wiener deconvolution is expressing a ``minimal energy'' constraint on
the deconvolved signal, and has been used in the framework of
psychophysiological interaction analysis to compute the interaction between a
seed's activity-inducing timecourse and an experimental modulation
\citep{Glover1999DeconvolutionImpulseResponse,Gitelman2003Modelingregionalpsychophysiologic,
Gerchen2014Analyzingtaskdependent,Di2018TaskConnectomicsExamining,
Freitas2020Timeresolvedeffective}.
Complementarily, the interest in deconvolution has increased to explore
time-varying activity in resting-state fMRI data
\citep{Preti2017dynamicfunctionalconnectome,Keilholz2017TimeResolvedResting,
Lurie2020Questionscontroversiesstudy,Bolton2020TappingMultiFaceted}.
In that case, the aim is to gain better insights of the neural signals that
drive functional connectivity at short time scales, as well as learning about
the spatio-temporal structure of functional components that dynamically
construct resting-state networks and their interactions
\citep{Karahanoglu2017Dynamicslargescale}.

Deconvolution of the resting-state fMRI signal has illustrated the significance
of transient, sparse spontaneous events
\citep{Petridou2012PeriodsrestfMRI,Allan2015FunctionalConnectivityMRI} that
refine the hierarchical clusterization of functional networks
\citep{Karahanoglu2013TotalactivationfMRI} and reveal their temporal overlap
based on their signal innovations not only in the human brain
\citep{Karahanoglu2015Transientbrainactivity}, but also in the spinal cord
\citep{Kinany2020DynamicFunctionalConnectivity}. Similar to task-related
studies, deconvolution allows to investigate modulatory interactions within and
between resting-state functional networks
\citep{Di2013ModulatoryInteractionsResting,Di2015Characterizationsrestingstate}.
In addition, decoding of the deconvolved spontaneous events allows to decipher
the flow of spontaneous thoughts and actions across different cognitive and
sensory domains while at rest
\citep{Karahanoglu2015Transientbrainactivity,GonzalezCastillo2019Imagingspontaneousflow,Tan_2017}.
Beyond findings on healthy subjects, deconvolution techniques have also proven
its utility in clinical conditions to characterize functional alterations of
patients with a progressive stage of multiple sclerosis at rest
\citep{Bommarito2020Alteredanteriordefault}, to find functional signatures
of prodromal psychotic symptoms and anxiety at rest on patients suffering from
schizophrenia \citep{Zoeller2019Largescalebrain}, to detect the foci of
interictal events in epilepsy patients without an EEG recording
\citep{Lopes2012Detectionepilepticactivity,Karahanoglu2013Spatialmappinginterictal},
or to study functional dissociations observed during non-rapid eye movement
sleep that are associated with reduced consolidation of information and impaired
consciousness \citep{Tarun2020NREMsleepstages}.

The algorithms for hemodynamic deconvolution can be classified based on the
assumed hemodynamic model and the optimization problem used to estimate the
neuronal-related signal. Most approaches assume a linear time-invariant model
for the hemodynamic response that is inverted by means of variational
(regularized) least squares estimators
\citep{Glover1999DeconvolutionImpulseResponse,Gitelman2003Modelingregionalpsychophysiologic,
Gaudes2010Detectioncharacterizationsingle,Gaudes2012Structuredsparsedeconvolution,
Gaudes2013Paradigmfreemapping,CaballeroGaudes2019deconvolutionalgorithmmulti,
HernandezGarcia2011Neuronaleventdetection,Karahanoglu2013TotalactivationfMRI,
Cherkaoui2019SparsitybasedBlind,
Huetel2021Hemodynamicmatrixfactorization,Costantini2022Anisotropic4DFiltering},
logistic functions
\citep{Bush2013Decodingneuralevents,Bush2015deconvolutionbasedapproach,
Loula2018DecodingfMRIactivity}, probabilistic mixture models
\citep{Pidnebesna2019EstimatingSparseNeuronal}, convolutional autoencoders
\citep{Huetel2018NeuralActivationEstimation} or nonparametric homomorphic
filtering \citep{Sreenivasan2015NonparametricHemodynamicDeconvolution}.
Alternatively, several methods have also been proposed to invert non-linear
models of the neuronal and hemodynamic coupling
\citep{Riera2004statespacemodel,Penny2005Bilineardynamicalsystems,Friston2008DEMvariationaltreatment,
Havlicek2011Dynamicmodelingneuronal,Aslan2016Jointstateparameter,
Madi2017HybridCubatureKalman,RuizEuler2018NonlinearDeconvolutionSampling}.

Among the variety of approaches, those based on regularized least squares
estimators have been employed more often due to their appropriate performance at
small spatial scales (e.g., voxelwise). Relevant for this work, two different
formulations can be established for the regularized least-squares deconvolution
problem, either based on a synthesis- or analysis-based model
\citep{Elad2007Analysisversussynthesis,ortelli2019synthesis}. The rationale
of the synthesis-based model is that we know or suspect that the true signal
(here, the neuronally-driven BOLD component of the fMRI signal) can be
represented as a linear combination of predefined patterns or dictionary atoms
(for instance, the hemodynamic response function). In contrast, the
analysis-based approach considers that the true signal is analyzed by some
relevant operator and the resulting signal is small (i.e., sparse).

As members of the groups that developed Paradigm Free Mapping (synthesis-based
solved with regularized least-squares estimators such as
ridge-regression \citealt{Gaudes2010Detectioncharacterizationsingle} or LASSO
\citealt{Gaudes2013Paradigmfreemapping}) and Total Activation (analysis-based
also solved with a regularized least-squares estimator using
generalized total variation
\citealt{Karahanoglu2011SignalProcessingApproach,Karahanoglu2013TotalactivationfMRI}
) deconvolution methods for fMRI data analysis, we are often contacted by
researchers who want to know about the similarities and differences between the
two methods and which one is better. \emph{It depends}---and to clarify this
point, this note revisits synthesis- and analysis-based deconvolution methods
for fMRI data and comprises four sections. First, we present the theory behind
these two deconvolution approaches based on regularized least squares estimators
that promote sparsity: Paradigm Free Mapping (PFM)
\citep{Gaudes2013Paradigmfreemapping} --- available in AFNI as
\textit{3dPFM}\footnote{\url{https://afni.nimh.nih.gov/pub/dist/doc/program_help/3dPFM.html}}
and
\textit{3dMEPFM}\footnote{\url{https://afni.nimh.nih.gov/pub/dist/doc/program_help/3dMEPFM.html}}
for single-echo and multi-echo data, respectively --- and Total Activation (TA)
\citep{Karahanoglu2013TotalactivationfMRI} --- available as part of the
\textit{iCAPs toolbox}\footnote{\url{https://c4science.ch/source/iCAPs/}}. We
describe the similarities and differences in their analytical formulations, and
how they can be related to each other. Next, we assess their performance
controlling for a fair comparison on simulated and experimental data. Finally,
we discuss their benefits and shortcomings and conclude with our vision on
potential extensions and developments.

%% file: sections/theory.tex

\section{Theory}

\subsection{Notations and definitions}

Matrices of size $N$ rows and $M$ columns are denoted by boldface capital
letters, e.g., $\mathbf{X} \in \mathbb{R}^{N\times M}$, whereas column vectors
of length $N$ are denoted as boldface lowercase letters, e.g., $\mathbf{x} \in
\mathbb{R}^{N}$. Scalars are denoted by lowercase letters, e.g., $k$. Continuous
functions are denoted by brackets, e.g., $h(t)$, while discrete functions are
denoted by square brackets, e.g., $x[k]$. The Euclidean norm of a matrix
$\mathbf{X}$ is denoted as $\|\mathbf{X}\|_2$, the $\ell_1$-norm is denoted by
$\| \mathbf{X} \|_1$ and the Frobenius norm is denoted by $\| \mathbf{X} \|_F$.
The discrete integration ($\mathbf{L}$) and difference ($\mathbf{D}$) operators
are defined as:

$$
\mathbf{L} = \left[\begin{array}{ccccc}
1 & 0 & \ldots & & \\
1 & 1 & 0 & \ldots & \\
1 & 1 & 1 & 0 & \ldots \\
\vdots & \ddots & \ddots & \ddots & \ddots
\end{array}\right], \quad \mathbf{D} = \left[\begin{array}{ccccc}
1 & 0 & \ldots & & \\
1 & -1 & 0 & \ldots & \\
0 & \ddots & \ddots & \ddots & \ldots \\
\vdots & \ddots & 0 & 1 & -1
\end{array}\right].
$$

\subsection{Conventional general linear model analysis}

Conventional general linear model (GLM) analysis puts forward a number of
regressors incorporating the knowledge about the paradigm or behavior. For
instance, the timing of epochs for a certain condition can be modeled as an
indicator function $p(t)$ (e.g., Dirac functions for event-related designs or
box-car functions for block-designs) convolved with the hemodynamic response
function (HRF) $h(t)$, and sampled at TR resolution
\citep{Friston1994AnalysisfunctionalMRI,Friston1998EventRelatedfMRI,
Boynton1996LinearSystemsAnalysis,Cohen1997ParametricAnalysisfMRI}:
$$
   x(t) = p*h(t) \rightarrow x[k] = p*h(k\cdot\text{TR}).
$$
The vector $\mathbf{x}=[x[k]]_{k=1,\ldots,N} \in \mathbb{R}^{N}$ then
constitutes the regressor modelling the hypothetical response, and several of
them can be stacked as columns of the design matrix $\mathbf{X}=[\mathbf{x}_1
\ldots \mathbf{x}_L] \in \mathbb{R}^{N \times L}$, leading to the well-known GLM
formulation: 
\begin{equation}
    \label{eq:glm}
    \mathbf{y} = \mathbf{X} \boldsymbol\beta + \mathbf{e},
\end{equation}
where the empirical timecourse $\mathbf{y} \in \mathbb{R}^{N}$ is explained by a
linear combination of the regressors in $\mathbf{X}$ weighted by the parameters
in $\boldsymbol\beta \in \mathbb{R}^{L}$ and corrupted by additive noise
$\mathbf{e}\in \mathbb{R}^{N}$. Under independent and identically distributed
Gaussian assumptions of the latter, the maximum likelihood estimate of the
parameter weights reverts to the ordinary least-squares estimator; i.e.,
minimizing the residual sum of squares between the fitted model and
measurements. The number of regressors $L$ is typically much less than the
number of measurements $N$, and thus the regression problem is over-determined
and does not require additional constraints or assumptions
\citep{HENSON2007178}.

In the deconvolution approach, no prior knowledge of the hypothetical response
is taken into account, and the purpose is to estimate the deconvolved
activity-inducing signal $\mathbf{s}$ from the measurements $\mathbf{y}$, which
can be formulated as the signal model
\begin{equation}
    \label{eq:synthesis_model}
    \mathbf{y} = \mathbf{Hs} + \mathbf{e},
\end{equation}
where $\mathbf{H} \in \mathbb{R}^{N \times N}$ is a Toeplitz matrix that
represents the discrete convolution with the HRF, and $\mathbf{s} \in
\mathbb{R}^{N}$ is a length-$N$ vector with the unknown activity-inducing
signal. Note that the temporal resolution of the activity-inducing signal and
the corresponding Toeplitz matrix is generally assumed to be equal to the TR of
the acquisition, but it could also be higher if an upsampled estimate is
desired. Despite the apparent similarity with the GLM equation, there are two
important differences. First, the multiplication with the design matrix of the
GLM is an expansion as a weighted linear combination of its columns, while the
multiplication with the HRF matrix represents a convolution operator. Second,
determining $\mathbf{s}$ is an ill-posed problem given the nature of the HRF. As
it can be seen intuitively, the convolution matrix $\mathbf{H}$ is highly
collinear (i.e., its columns are highly correlated) due to large overlap between
shifted HRFs (see Figure \ref{fig:sim_and_hrf}C), thus introducing uncertainty
in the estimates of $\mathbf{s}$ when noise is present. Consequently, additional
assumptions under the form of regularization terms (or priors) in the estimate
are needed to reduce their variance. In the least squares sense, the
optimization problem to solve is given by 
\begin{equation}
    \label{eq:regularized_least_squares}
    \hat{\mathbf{s}} = \arg \min_{\mathbf{s}} \frac{1}{2} \| \mathbf{y} - \mathbf{Hs} \|_2^2 + \Omega(\mathbf{s}).
\end{equation}
The first term quantifies data fitness, which can be justified as the
log-likelihood term derived from Gaussian noise assumptions, while the second
term \(\Omega(\mathbf{s})\) brings in regularization and can be interpreted as a
prior on the activity-inducing signal. For example, the $\ell_2$-norm of
$\mathbf{s}$ (i.e., $\Omega(\mathbf{s})=\lambda \left\| \mathbf{s}\right\|_2^2$)
is imposed for ridge regression or Wiener deconvolution, which introduces a
trade-off between the data fit term and ``energy'' of the estimates that is
controlled by the regularization parameter $\lambda$. 
regularized terms are related to the elastic net (i.e.,
$\Omega(\mathbf{x})=\lambda_1\|\mathbf{x}\|_2^2 + \lambda_2\|\mathbf{x}\|_1$)
[REF]. 

\subsection{Paradigm Free Mapping}
 In paradigm free mapping (PFM), the formulation of
 Eq.~(\ref{eq:regularized_least_squares}) was considered equivalently as fitting
 the measurements using the atoms of the HRF dictionary (i.e., columns of
 $\mathbf{H}$) with corresponding weights (entries of $\mathbf{s}$). This model
 corresponds to a synthesis formulation. In
 \citealt{Gaudes2013Paradigmfreemapping} a sparsity-pursuing regularization term
 was introduced on $\mathbf{s}$, which in a strict way reverts to choosing
 \(\Omega(\mathbf{s})=\lambda \| \mathbf{s} \|_0\) as the regularization term
 and solving the optimization problem
 \citep{Bruckstein2009SparseSolutionsSystems}. However, finding the optimal
 solution to the problem demands an exhaustive search across all possible
 combinations of the columns of \(\mathbf{H}\). Hence, a  pragmatic solution is
 to solve the convex-relaxed optimization problem for the \(l_1\)-norm, commonly
 known as Basis Pursuit Denoising \citep{Chen2001BasisPursuitDenoising} or
 equivalently as the least absolute shrinkage and selection operator (LASSO)
 \citep{Tibshirani1996RegressionShrinkageSelection}: 
\begin{equation}
    \label{eq:pfm_spike}
    \hat{\mathbf{s}} = \arg \min_{\mathbf{s}} \frac{1}{2} \| \mathbf{y} - \mathbf{Hs} \|_2^2 + \lambda \| \mathbf{s} \|_1,
\end{equation}
which provides fast convergence to a global solution. Imposing sparsity on the
activity-inducing signal implies that it is assumed to be well represented by a
reduced subset of few non-zero coefficients at the fMRI timescale, which in turn
trigger event-related BOLD responses. Hereinafter, we refer to this assumption
as the \textit{spike model}. However, even if PFM was developed
as a spike model, its formulation in Eq.\eqref{eq:pfm_spike} can be extended to
estimate the innovation signal, i.e., the derivative of the activity-inducing
signal, as shown in section~\ref{sec:unifying_both_perspectives}.


\subsection{Total Activation}
Alternatively, deconvolution can be formulated as if the signal to be recovered
directly fits the measurements and at the same time satisfies some suitable
regularization, which leads to
\begin{equation}
\label{eq:analysis_model}
    \hat{\mathbf{x}} = \arg \min_{\mathbf{x}} \frac{1}{2} \| \mathbf{y} - \mathbf{x} \|_2^2 + \Omega(\mathbf{x}).
\end{equation}
Under this analysis formulation, total variation (TV), i.e., the $\ell_1$-norm
of the derivative $\Omega(\mathbf{x})=\lambda \|\mathbf{Dx}\|_1$, is a powerful
regularizer since it favors recovery of piecewise-constant signals
\citep{Chambolle2004TotalVariation}. Going beyond, the approach of generalized
TV introduces an additional differential operator $\mathbf{D_H}$ in the
regularizer that can be tailored as the inverse operator of a linear
system~\citep{Karahanoglu2011SignalProcessingApproach}, that is,
$\Omega(\mathbf{x})=\lambda \|\mathbf{D D_H x}\|_1$. In the context of
hemodynamic deconvolution, Total Activation is proposed for which the discrete
operator $\mathbf{D_H}$ is derived from the inverse of the continuous-domain
linearized Balloon-Windkessel model
\citep{Buxton1998BalloonModel,Friston2000Nonlinear-Balloon}. 
The interested reader is referred to
\citep{Khalidov2011ActiveletsWaveletssparse,Karahanoglu2011SignalProcessingApproach,
Karahanoglu2013TotalactivationfMRI} for a detailed description of this
derivation. 

Therefore, the solution of the Total Activation (TA) problem
\begin{equation}
\label{eq:TA}
    \hat{\mathbf{x}} = \arg \min_{\mathbf{x}} \frac{1}{2} \| \mathbf{y} - \mathbf{x} \|_2^2 + \lambda \|\mathbf{D D_H x} \|_1
\end{equation}
will yield the activity-related signal $\mathbf{x}$ for which the
activity-inducing signal $\mathbf{s}=\mathbf{D_H x}$ and the so-called
innovation signal $\mathbf{u}=\mathbf{Ds}$, i.e., the derivate
of the activity-inducing signal, will also be available, as they are required
for the regularization. We refer to modeling the activity-inducing signal based
on the innovation signal as the \textit{block model}.
Nevertheless, even if TA was originally developed as a block
model, its formulation in Eq.\eqref{eq:TA} can be made equivalent to the spike
model as shown in section~\ref{sec:unifying_both_perspectives}.

\subsection{Unifying both perspectives}
\label{sec:unifying_both_perspectives}
PFM and TA are based on the synthesis- and analysis-based formulation of the
deconvolution problem, respectively. They are also tailored for the spike and
block model, respectively. In the first case, the recovered deconvolved signal
is synthesized to be matched to the measurements, while in the second case, the
recovered signal is directly matched to the measurements but needs to satisfy
its analysis in terms of deconvolution. This also corresponds to using the
forward or backward model of the hemodynamic system, respectively. Hence, it is
possible to make both approaches equivalent
\citep{Elad2007Analysisversussynthesis}\footnote{Without dwelling into
technicalities, for total variation, this equivalence is correct up to the
constant, which is in the null space of the derivative operator.}.

To start with, TA can be made equivalent to PFM by adapting it for the spike
model; i.e., when removing the derivative operator $\mathbf{D}$ of the
regularizer in Eq. (\ref{eq:TA}), it can be readily verified that replacing in
that case $\mathbf{x}=\mathbf{Hs}$ leads to identical equations and thus both
assume a spike model, since $\mathbf{H}$ and $\mathbf{D_H}$ will cancel out each
other \citep{Karahanoglu2011SignalProcessingApproach}\footnote{Again, this holds
up to elements of the null space.}.

Conversely, the PFM spike model can also accommodate the
\sout{TA} block model by modifying Eq. (\ref{eq:pfm_spike})
with the forward model $\mathbf{y} = \mathbf{H L u} + \mathbf{e}$. Here, the
activity-inducing signal $\mathbf{s}$ is rewritten in terms of the innovation
signal $\mathbf{u}$ as $\mathbf{s}=\mathbf{Lu}$ where the matrix $\mathbf{L}$ is
the first-order integration operator
\citep{Cherkaoui2019SparsitybasedBlind,Urunuela2020StabilityBasedSparse}. This
way, PFM can estimate the innovation signal $\mathbf{u}$ as follows: 
\begin{equation}
    \label{eq:pfm_block}
    \hat{\mathbf{u}} = \arg \min_{\mathbf{u}} \frac{1}{2} \| \mathbf{y} - \mathbf{HLu} \|_2^2 + \lambda \| \mathbf{u} \|_1,
\end{equation}
and becomes equivalent to TA by replacing $\mathbf{u}=\mathbf{D D_H x}$, and
thus adopting the block model. Based on the previous equations
(\ref{eq:pfm_spike}), (\ref{eq:TA}) and (\ref{eq:pfm_block}), it is clear that
both PFM and TA can operate under the spike and block models, providing a
convenient signal model according to the different assumptions of the underlying
neuronal-related signal. This work evaluates the core of the two techniques;
i.e., the regularized least-squares problem with temporal regularization without
considering the spatial regularization term originally incorporated in TA. For
the remainder of this paper, we will use the PFM and TA formalisms with both
spike and block models. 

\begin{figure}[t!]
    \begin{center}
        \includegraphics[width=\columnwidth]{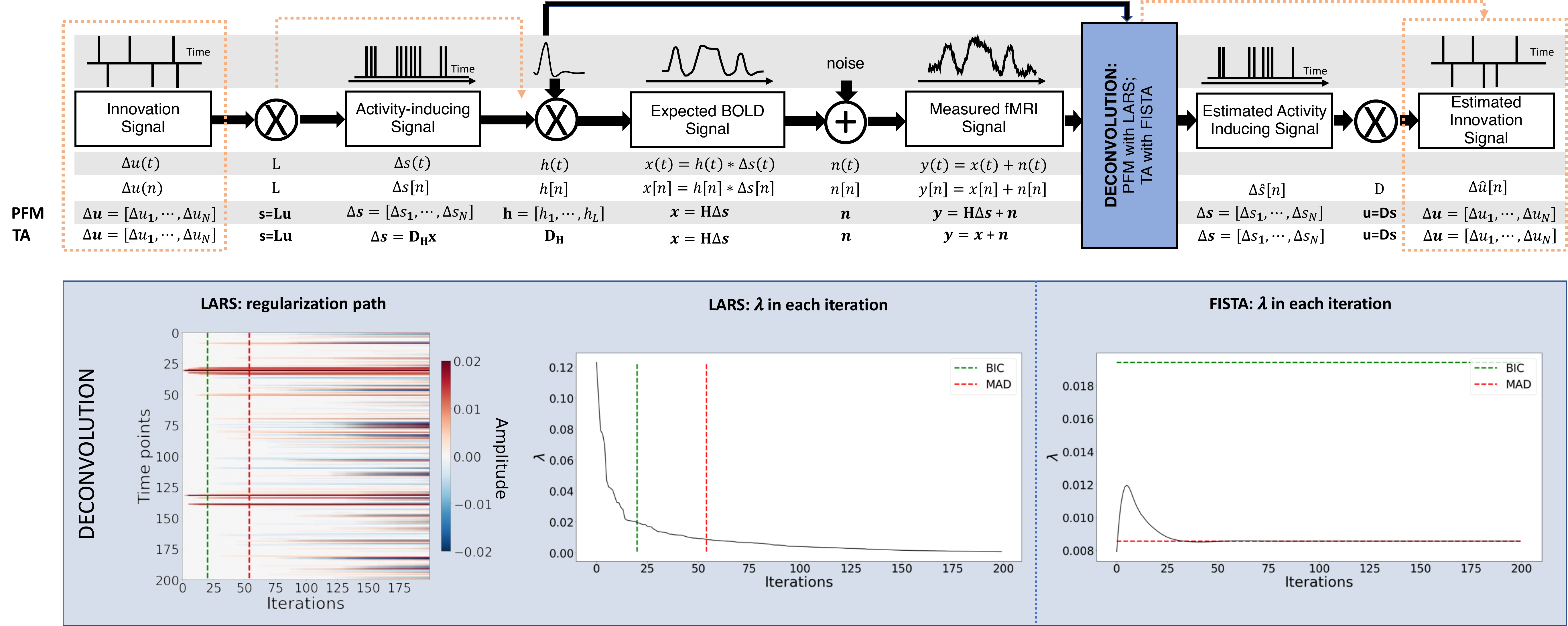}
    \end{center}
    \caption{Flowchart detailing the different steps of the fMRI signal and the
    deconvolution methods described. The orange arrows indicate the flow to
    estimate the innovation signals, i.e., the derivative of
    the activity-inducing signal. The blue box depicts the iterative
    \textit{modus operandi} of the two algorithms used in this paper to solve
    the paradigm free mapping (PFM) and total activation (TA)
    deconvolution problems. The plot on the left shows the regularization path
    obtained with the least angle regression (LARS) algorithm,
    where the x-axis illustrates the different iterations of the algorithm, the
    y-axis represents points in time, and the color describes the amplitude of
    the estimated signal. The middle plot depicts the decreasing values of
    $\lambda$ for each iteration of LARS as the regularization path is computed.
    The green and r ed dashed lines in both plots illustrate the
    Bayesian information criterion (BIC) and
    median absolute deviation (MAD) solutions, respectively.
    Comparatively, the changes in $\lambda$ when the
    fast iterative shrinkage-thresholding algorithm (FISTA)
    method is made to converge to the MAD estimate of the noise are shown on the
    right. Likewise, the $\lambda$ corresponding to the BIC and MAD solutions
    are shown with dashed lines.}
\label{fig:flowchart}
\end{figure}

\subsection{Algorithms and parameter selection}
\label{sec:regparam}
Despite their apparent resemblance, the practical implementations of the PFM and
TA methods proposed different algorithms to solve the corresponding optimization
problem and select an adequate regularization parameter $\lambda$
\citep{Gaudes2013Paradigmfreemapping,Karahanoglu2013TotalactivationfMRI}. The
PFM implementation available in AFNI employs the least angle regression (LARS)
\citep{Efron2004Leastangleregression}, whereas the TA implementation uses the
fast iterative shrinkage-thresholding algorithm (FISTA)
\citep{Beck2009FastIterativeShrinkage}. The blue box in Figure
\ref{fig:flowchart} provides a descriptive view of the iterative \textit{modus
operandi} of the two algorithms.

On the one hand, LARS is a homotopy approach that computes all the possible
solutions to the optimization problem and their corresponding value of
$\lambda$; i.e., the regularization path, and the solution according to the
Bayesian Information Criterion (BIC)
\citep{Schwarz1978EstimatingDimensionModel}, was recommended as the most
appropriate in the case of PFM approaches since AIC often tends to overfit the
signal\citep{Gaudes2013Paradigmfreemapping,CaballeroGaudes2019deconvolutionalgorithmmulti}.

On the other hand, FISTA is an extension of the classical gradient algorithm
that provides fast convergence for large-scale problems. In the case of FISTA
though, the regularization parameter $\lambda$ must be selected prior to solving
the problem, but can be updated in every iteration so that the residuals of the
data fit converge to an estimated noise level of the data $\hat{\sigma}$:
\begin{equation}
    \lambda^{n+1} = \frac{N \hat{\sigma}}{\frac{1}{2} \| \mathbf{y} - \mathbf{x}^n \|_F^2} \lambda^n,
\label{eq:std}
\end{equation}
where $x^n$ is the $n^{th}$ iteration estimate, $\lambda^n$ and $\lambda^{n+1}$
are the $n^{th}$ and $n+1^{th}$ iteration values for the regularization
parameter $\lambda$, and $N$ is the number of points in the time-course. The
pre-estimated noise level can be obtained as the median absolute deviation (MAD)
of the fine-scale wavelet coefficients (Daubechies, order 3) of the fMRI
timecourse. The MAD criterion has been adopted in TA
\citep{Karahanoglu2013TotalactivationfMRI}. Of note, similar formulations based
on the MAD estimate have also been applied in PFM formulations
\citep{Gaudes2012Structuredsparsedeconvolution,Gaudes2011MorphologicalPFM}.


%% file: sections/methods.tex

\section{Methods}
\label{sec:data}

\subsection{Simulated data}

\begin{figure}[t!]
    \begin{center}
        \includegraphics[width=0.75\columnwidth]{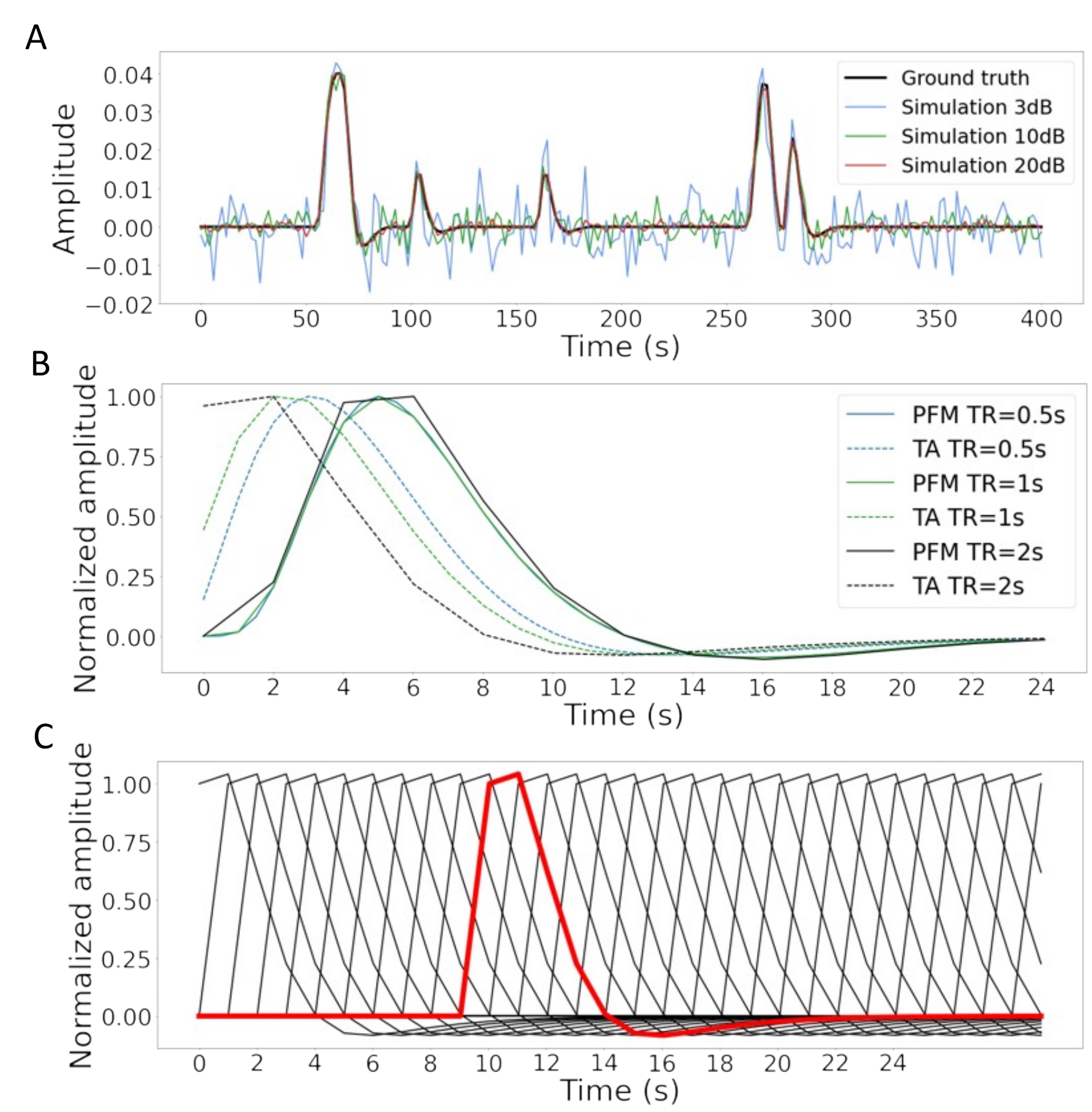}
    \end{center}
    \caption{A) Simulated signal with different SNRs (20 dB, 10 dB and 3 dB) and ground truth given in signal percentage change (SPC). B) Canonical HRF models typically used by PFM (solid line) and TA (dashed line) at TR = 0.5 s (blue), TR = 1 s (green) and TR = 2 s (black). Without loss of generality, the waveforms are scaled to unit amplitude for visualization. C) Representation of shifted HRFs at TR = 2 s that build the design matrix for PFM when the HRF model has been matched to that in TA. The red line corresponds to one of the columns of the HRF matrix.}
\label{fig:sim_and_hrf}
\end{figure}

In order to compare the two methods while controlling for their correct
performance, we created a simulation scenario that can be found in the GitHub
repository shared in Section~\ref{sec:github}. For the sake of illustration, we
describe here the simulations corresponding to a timecourse with a duration of
400 seconds (TR = 2 s) where the activity-inducing signal includes 5 events,
which are convolved with the canonical HRF. Different noise sources
(physiological, thermal, and motion-related) were also added and we simulated
three different scenarios with varying signal-to-noise ratios (SNR = [20 dB, 10
dB, 3 dB]) that represent high, medium and low contrast-to-noise ratios as shown
in Figure~\ref{fig:sim_and_hrf}A. Noise was created following the procedure in
\citep{Gaudes2013Paradigmfreemapping} as the sum of uncorrelated Gaussian noise
and sinusoidal signals to simulate a realistic noise model with thermal noise,
cardiac and respiratory physiological fluctuations, respectively. The
physiological signals were generated as
\begin{equation}
    \sum_{i=1}^{2} \frac{1}{2^{i-1}}\left(\sin \left(2 \pi f_{r, i} t+\phi_{\mathrm{r}, i}\right)+\sin \left(2 \pi f_{c, i} t+\phi_{c, i}\right)\right),
\end{equation}
with up to second-order harmonics per cardiac (\(f_{c,i}\)) and respiratory
(\(f_{r,i}\)) component that were randomly generated following normal
distributions with variance 0.04 and mean \(if_r\) and \(if_c\), for \(i = [1,
2]\). We set the fundamental frequencies to \(f_r = 0.3\) Hz for the respiratory
component \citep{Birn2006Separatingrespiratoryvariation}) and \(f_c = 1.1\) Hz
for the cardiac component \citep{Shmueli2007Lowfrequencyfluctuations}). The
phases of each harmonic \(\phi\) were randomly selected from a uniform
distribution between 0 and 2$\pi$ radians. To simulate physiological noise that
is proportional to the change in BOLD signal, a variable ratio between the
physiological (\(\sigma_P\)) and the thermal (\(\sigma_0\)) noise was modeled as
\(\sigma_P/\sigma_0 = a(tSNR)^b + c\), where \(a = 5.01 \times 10^{-6}\), \(b =
2.81\), and \(c = 0.397\), following the experimental measures available in
Table 3 from \citep{Triantafyllou2005Comparisonphysiologicalnoise}).

\subsection{Experimental data}
To compare the performance of the two approaches as well as illustrate their
operation, we employ two representative experimental datasets.

\textbf{Motor task dataset:} One healthy subject was scanned in a 3T MR scanner
(Siemens) under a Basque Center on Cognition, Brain and Language Review
Board-approved protocol. T2*-weighted multi-echo fMRI data was acquired with a
simultaneous-multislice multi-echo gradient echo-planar imaging sequence, kindly
provided by the Center of Magnetic Resonance Research (University of Minnesota,
USA) \citep{Feinberg_2010,Moeller_2010,Setsompop_2011}, with the following
parameters: 340 time frames, 52 slices, Partial-Fourier =
6/8, voxel size = $2.4\times2.4\times3$ mm\textsuperscript{3}, TR = 1.5 s, TEs =
10.6/28.69/46.78/64.87/82.96 ms, flip angle = 70\(^o\), multiband factor = 4,
GRAPPA = 2.

During the fMRI acquisition, the subject performed a motor task consisting of
five different movements (left-hand finger tapping, right-hand finger tapping,
moving the left toes, moving the right toes and moving the tongue) that were
visually cued through a mirror located on the head coil. These conditions were
randomly intermixed every 16 seconds, and were only repeated once the entire set
of stimuli were presented. Data preprocessing consisted of first, discarding the
first 10 volumes of the functional data to achieve a steady state of
magnetization. Then, image realignment to the skull-stripped single-band
reference image (SBRef) was computed on the first echo, and the estimated
rigid-body spatial transformation was applied to all other echoes
\citep{Jenkinson2012FSL,Jenkinson_2001}. A brain mask obtained from the SBRef
volume was applied to all the echoes and the different echo timeseries were
optimally combined (OC) voxelwise by weighting each timeseries contribution by
its T2* value \citep{Posse_1999}. AFNI \citep{Cox1996AFNISoftwareAnalysis} was
employed for a detrending of up to 4\textsuperscript{th}-order Legendre
polynomials, within-brain spatial smoothing (3 mm FWHM) and voxelwise signal
normalization to percentage change. Finally, distortion field correction was
performed on the OC volume with Topup \citep{Andersson_2003}, using the pair of
spin-echo EPI images with reversed phase encoding acquired before the ME-EPI
acquisition \citep{Glasser_2016}.

\textbf{Resting-state datasets:} One healthy subject was scanned in a 3T MR
scanner (Siemens) under a Basque Center on Cognition, Brain and Language Review
Board-approved protocol. Two runs of T2*-weighted fMRI data were acquired during
resting-state, each with 10 min duration, with 1) a standard gradient-echo
echo-planar imaging sequence (monoband) (TR = 2000 ms, TE = 29 ms, flip-angle =
78\(^o\), matrix size = $64\times64$, voxel size = $3\times3\times3$
mm\textsuperscript{3}, 33 axial slices with interleaved acquisition, slice gap =
0.6 mm) and 2) a  simultaneous-multislice gradient-echo echo-planar imaging
sequence (multiband factor = 3, TR = 800 ms, TE = 29 ms, flip-angle = 60\(^o\),
matrix size = $64\times64$, voxel size = $3\times3\times3$
mm\textsuperscript{3}, 42 axial slices with interleaved acquisition, no slice
gap). Single-band reference images were also collected in both resting-state
acquisitions for head motion realignment. Field maps were also obtained to
correct for field distortions.

During both acquisitions, participants were instructed to keep their eyes open,
fixating a white cross that they saw through a mirror located on the head coil,
and not to think about anything specific. The data was pre-processed using AFNI
\citep{Cox1996AFNISoftwareAnalysis}. First, volumes corresponding to the initial
10 seconds were removed to allow for a steady-state magnetization. Then, the
voxel time-series were despiked to reduce large-amplitude deviations and
slice-time corrected. Inhomogeneities caused by magnetic susceptibility were
corrected with FUGUE (FSL) using the field map images \citep{Jenkinson2012FSL}.
Next, functional images were realigned to a base volume (monoband: volume with
the lowest head motion; multiband: single-band reference image). Finally, a
simultaneous nuisance regression step was performed comprising up to
6\textsuperscript{th}-order Legendre polynomials, low-pass filtering with a
cutoff frequency of 0.25 Hz (only on multiband data to match the frequency
content of the monoband), 6 realignment parameters plus temporal derivatives, 5
principal components of white matter (WM), 5 principal components of lateral
ventricle voxels (anatomical CompCor) \citep{Behzadi_2007} and 5 principal
components of the brain's edge voxels ,\citep{Patriat_2015}. WM, CSF and brain's
edge-voxel masks were obtained from Freesurfer tissue and brain segmentations.
In addition, scans with potential artifacts were identified and censored when
the euclidean norm of the temporal derivative of the realignment parameters
(ENORM) was larger than 0.4, and the proportion of voxels adjusted in the
despiking step exceeded 10\%.

\subsection{Selection of the hemodynamic response function}

In their original formulations, PFM and TA specify the discrete-time HRF in
different ways. For PFM, the continuous-domain specification of the canonical
double-gamma HRF \citep{HENSON2007178} is sampled at the TR and then put as
shifted impulse responses to build the matrix $\mathbf{H}$.  In the case of TA,
however, the continuous-domain linearized version of the balloon-windkessel
model is discretized to build the linear differential operator in
$\mathbf{D_H}$. While the TR only changes the resolution of the HRF shape for
PFM, the impact of an equivalent impulse response of the discretized
differential operator at different TR is more pronounced. As shown in
Figure~\ref{fig:sim_and_hrf}B, longer TR leads to equivalent impulse responses
of TA that are shifted in time, provoking a lack of the initial baseline and
rise of the response. We refer the reader to Figure {\ref{fig:hrf_differences}}
to see the differences in the estimation of the activity-inducing and innovation
signals when both methods use the HRF in their original formulation. To avoid
differences between PFM and TA based on their built-in HRF, we choose to build
the synthesis operator $\mathbf{H}$ with shifted versions of the HRF given by
the TA analysis operator (e.g., see Figure~\ref{fig:sim_and_hrf}C for the TR=2s
case).

\subsection{Selection of the regularization parameter}

We use the simulated data to compare the performance of the two deconvolution
algorithms with both BIC and MAD criteria to set the regularization parameter
$\lambda$ (see section \ref{sec:regparam}). We also evaluate if the algorithms
behave differently in terms of the estimation of the activity-inducing signal
$\mathbf{\hat{s}}$ using the spike model described in~\eqref{eq:pfm_spike} and
the block model based on the innovation signal $\mathbf{\hat{u}}$
in~\eqref{eq:pfm_block}.

For selection based on the BIC, LARS was initally performed with the PFM
deconvolution model to obtain the solution for every possible $\lambda$ in the
regularization path. Then, the values of $\lambda$ corresponding to the BIC
optimum were adopted to solve the TA deconvolution model by means of FISTA. 

For a selection based on the MAD estimate of the noise, we apply the temporal
regularization in its original form for TA, whereas for PFM the selected
$\lambda$ corresponds to the solution whose residuals have the closest standard
deviation to the estimated noise level of the data $\hat{\sigma}$.  

\subsection{Analyses in experimental fMRI data}

\textbf{Difference between approaches}: To assess the discrepancies between both
approaches when applied on experimental fMRI data, we calculate the square root
of the sum of squares of the differences (RSSD) between the activity-inducing
signals estimated with PFM and TA on the three experimental datasets as
\begin{equation}
    \text{RSSD} = \sqrt{\frac{1}{N} \sum_{k=1}^N (\hat{s}_\text{PFM}[k] - \hat{s}_\text{TA}[k])^2},
\end{equation}
where $N$ is the number of timepoints of the acquisition. The RSSD of the
innovation signals $\mathbf{\hat{u}}$ was computed equally.

\textbf{Task fMRI data}: In the analysis of the motor task data, we evaluate the
performance of PFM and TA in comparison with a conventional General Linear Model
analysis (\textit{3dDeconvolve} in AFNI) that takes advantage of the information
about the duration and onsets of the motor trials. Given the block design of the
motor task, we only make this comparison with the block model.

\textbf{Resting-state fMRI data}: We also illustrate the usefulness of
deconvolution approaches in the analysis of resting-state data where information
about the timings of neuronal-related BOLD activity cannot be predicted. Apart
from being able to explore individual maps of deconvolved activity (i.e.,
innovation signals, activity-inducing signals, or hemodynamic signals) at the
temporal resolution of the acquisition (or deconvolution), here we calculate the
average extreme points of the activity-inducing and innovation
maps (given that these examples do not have a sufficient number of scans to
perform a clustering step) and illustrate how popular approaches like
co-activation patterns
(CAPs)\citep{Tagliazucchi2012,Liu2018Coactivationpatterns} and innovation-driven
co-activation patterns (iCAPs) \citep{Karahanoglu2015Transientbrainactivity} can
be applied on the deconvolved signals to reveal patterns of coordinated brain
activity. To achieve this, we calculate the average time-series in a seed of 9
voxels located in the precuneus, supramarginal gyrus, and occipital gyri
independently, and solve the deconvolution problem to find the activity-inducing
and innovation signals in the seeds. We then apply a 95\textsuperscript{th}
percentile threshold and average the maps of the time-frames that survive the
threshold. Finally, we apply the same procedure to the original--- i.e.,
non-deconvolved--- signal in the seed and compare the results with the
widely-used seed correlation approach.

%% file: sections/results.tex

\section{Results}

\subsection{Performance based on the regularization parameter}
\label{sec:regpath}

\begin{figure}[t!]
    \begin{center}
        \includegraphics[width=\textwidth]{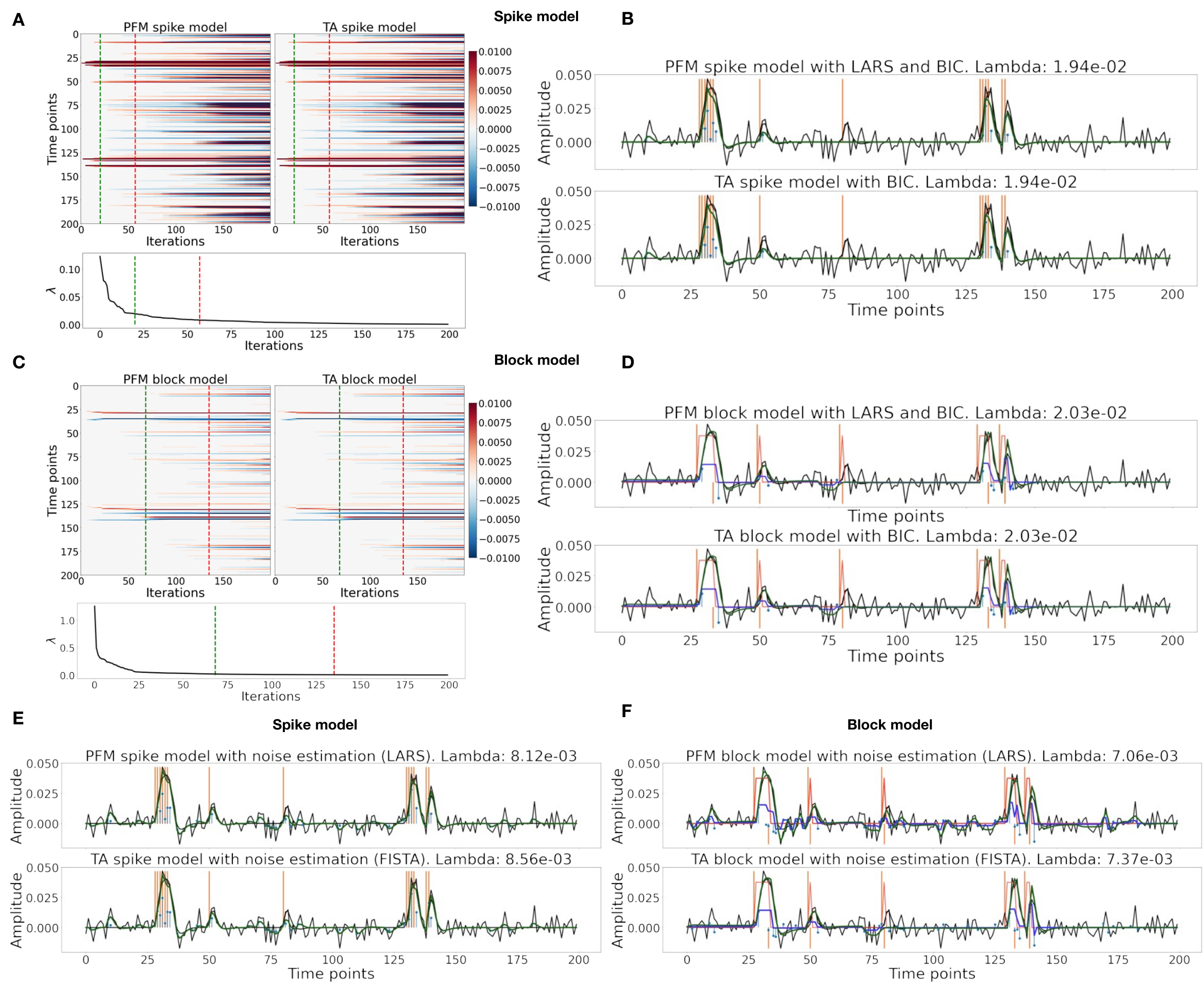}
    \end{center}
    \caption{(A) Heatmap of the regularization paths of the activity-inducing
    signals (spike model) estimated with PFM and TA as a function of $\lambda$
    for the simulated data with SNR = 3 dB (x-axis: increasing number of
    iterations or $\lambda$ as given by LARS; y-axis: time; color: amplitude).
    Vertical lines denote iterations corresponding to the BIC (dashed line) and
    MAD (dotted line) selection of $\lambda$. (B) Estimated activity-inducing
    (blue) and activity-related (green) signals with a selection of $\lambda$
    based on the BIC. Orange and red lines depict the ground truth. (C) Heatmap
    of the regularization paths of the innovation signals (block model)
    estimated with PFM and TA as a function of $\lambda$ for the simulated data
    with SNR = 3 dB. (D) Estimated innovation (blue), activity-inducing (darker
    blue), and activity-related (green) signals with a selection of $\lambda$
    based on the BIC. (E) Activity-inducing and activity-related (fit,
    $\mathbf{x}$) signals estimated with PFM (top) and TA (bottom) when
    $\lambda$ is selected based on the MAD method with the spike model, and (F)
    with the block model for the simulated data with SNR = 3 dB.}
\label{fig:sim}
\end{figure}

Figure~\ref{fig:sim}A shows the regularization paths of PFM and TA side by side
obtained for the spike model of Eq.~\eqref{eq:pfm_spike} for SNR=3 dB. The
solutions for all three SNR conditions are shown in Figures~\ref{fig:path_spike}
and \ref{fig:path_block}. Starting from the maximum $\lambda$ corresponding to a
null estimate and for decreasing values of $\lambda$, LARS computes a new
estimate at the value of $\lambda$ that reduces the sparsity promoted by the
\(l_1\)-norm and causes a change in the active set of non-zero coefficients of
the estimate (i.e., a zero coefficient becomes non-zero or vice versa) as shown
in the horizontal axis of the heatmaps. Vertical dashed lines depict the
selection of the regularization parameter based on the BIC, and thus, the
colored coefficients indicated by these depict the estimated activity-inducing
signal $\mathbf{\hat{{s}}}$. Figure~\ref{fig:sim}B illustrates the resulting
estimates of the activity-inducing and activity-related hemodynamic signals when
basing the selection of $\lambda$ on the BIC for SNR=3 dB. Given that the
regularization paths of both approaches are identical, it can be clearly
observed that the BIC-based estimates are identical too for the corresponding
$\lambda$. Thus, Figures~\ref{fig:sim}A, \ref{fig:sim}B, \ref{fig:path_spike}
and \ref{fig:path_block} demonstrate that, regardless of the simulated SNR
condition, the spike model of both deconvolution algorithms produces identical
regularization paths when the same HRF and regularization parameters are
applied, and hence, identical estimates of the activity-inducing signal
$\mathbf{\hat{{s}}}$ and neuronal-related hemodynamic signal
$\mathbf{\hat{{x}}}$. Likewise, Figure~\ref{fig:sim}C demonstrates that the
regularization paths for the block model defined in Eqs.~\eqref{eq:TA}
and~\eqref{eq:pfm_block} also yield virtually identical estimates of the
innovation signals for both PFM and TA methods. Again, the BIC-based selection
of $\lambda$ is identical for both PFM and TA. As illustrated in
Figure~\ref{fig:sim}D, the estimates of the innovation signal $\mathbf{u}$ also
show no distinguishable differences between the algorithms. 
Figures~\ref{fig:sim} A-D demonstrate that both PFM and TA yield equivalent
regularization paths and estimates of the innovation signal and
activity-inducing signal regardless of the simulated SNR condition when applying
the same HRF and regularization parameters with the block and spike models.

As for selecting $\lambda$ with the MAD criterion defined in Eq.~\eqref{eq:std},
Figure~\ref{fig:sim}E depicts the estimated activity-inducing and
activity-related signals for the simulated low-SNR setting using the spike
model, while Figure~\ref{fig:sim}F shows the estimated signals corresponding to
the block model. Both plots in Figure~\ref{fig:sim}E and F depict nearly
identical results between PFM and TA with both models. Given that the
regularization paths of both techniques are identical, minor dissimilarities are
owing to the slight differences in the selection of $\lambda$ due to the
quantization of the values returned by LARS.

\subsection{Performance on experimental data}

\begin{figure}[t!]
    \begin{center}
        \includegraphics[width=\textwidth]{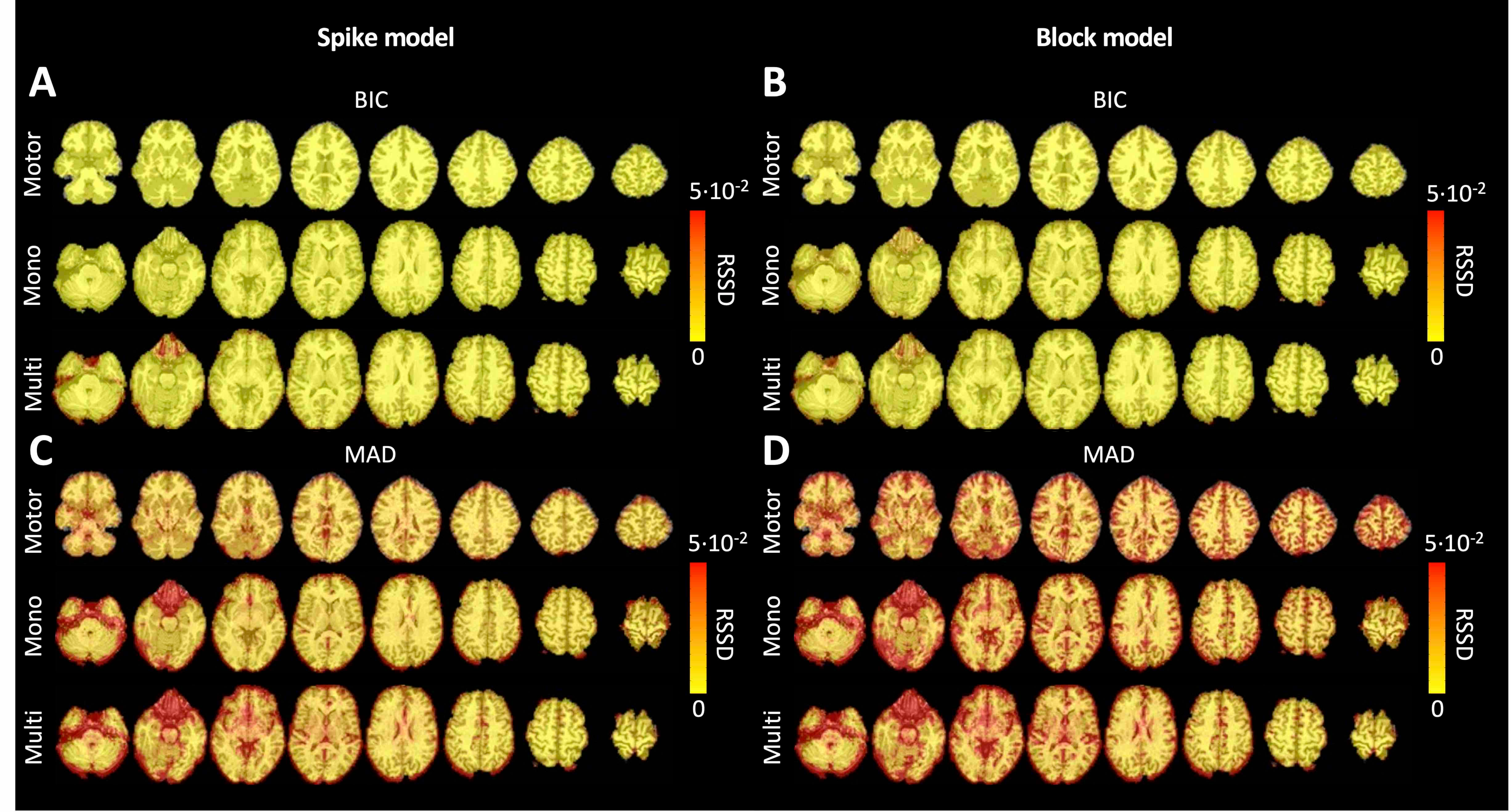}
    \end{center}
    \caption{Square root of the sum of squared differences (RSSD) between the estimates obtained with PFM and TA for (A) spike model (activity-inducing signal) and BIC selection of $\lambda$, (B) block model (innovation signal) and BIC selection, (C) spike model (activity-inducing signal) and MAD selection, (D) block model (innovation signal) and MAD selection. RSSD maps are shown for the three experimental fMRI datasets: the motor task (Motor), the monoband resting-state (Mono), and the multiband resting-state (Multi) datasets.}
\label{fig:rss}
\end{figure}

Figure~\ref{fig:rss} depicts the RSSD maps revealing differences between PFM and
TA estimates for the spike (Figure~\ref{fig:rss}A and C) and block
(Figure~\ref{fig:rss}B and D) models when applied to the three experimental fMRI
datasets. The RSSD values are virtually negligible (i.e., depicted in yellow) in
most of the within-brain voxels and lower than the amplitude of the estimates of
the activity-inducing and innovation signals. Based on the maximum value of the
range shown in each image, we observe that the similarity between both
approaches is more evident for the spike model (with both selection criteria)
and the block model with the BIC selection. However, given the different
approaches used for the selection of the regularization parameter $\lambda$
based on the MAD estimate of the noise (i.e., converging so that the residuals
of FISTA are equal to the MAD estimate of the noise for TA vs. finding the LARS
residual that is closest to the MAD estimate of the noise), higher RSSD values
can be observed with the largest differences occurring in gray matter voxels.
These areas also correspond to low values of $\lambda$ (see
Figure~\ref{fig:lambdas}) and MAD estimates of the noise (see
Figure~\ref{fig:mad_estimate}), while the highest values are visible in regions
with signal droupouts, ventricles, and white matter. These differences that
arise from the differing methods to find the optimal regularization parameter
based on the MAD estimate of the noise can be clearly seen in the root sum of
squares (RSS) of the estimates of the two methods
(Figure~\ref{fig:rss_comparison}). These differences are also observable in the
ATS calculated from estimates obtained with the MAD selection as shown in
Figure~\ref{fig:task_mad}. However, the identical regularization paths shown in
Figure~\ref{fig:motor_regpaths} demonstrate that both methods perform
equivalently on experimental data (see estimates of innovation signal obtained
with an identical selection of $\lambda$ in Figure~\ref{fig:mad_inno_ts}).
Hence, the higher RSSD values originate from the different methods to find the
optimal regularization parameter based on the MAD estimate of the noise that
yield differnt solutions as shown by the dashed vertical lines in
Figure~\ref{fig:motor_regpaths}.

\begin{figure}[t!]
    \begin{center}
        \includegraphics[width=\textwidth]{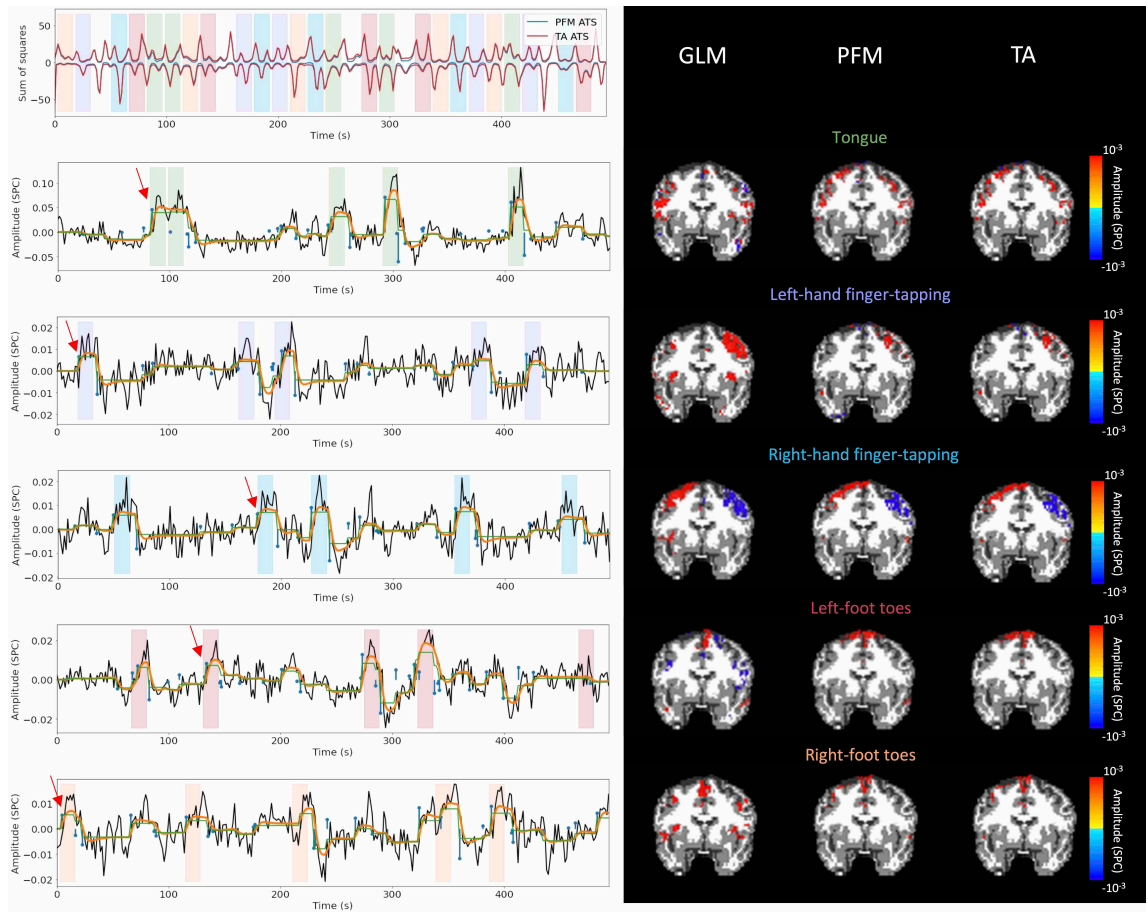}
    \end{center}
    \caption{Activity maps of the motor task using a selection of $\lambda$ based on the BIC estimate. Row 1: Activation time-series (ATS) of the innovation signals estimated by PFM (in blue) or TA (in red) calculated as the sum of squares of all voxels at every timepoint. Positive-valued and negative-valued contributions were separated into two distinct time-courses. Color-bands indicate the onset and duration of each condition in the task (green: tongue motion, purple: left-hand finger-tapping, blue: right-hand finger-tapping, red: left-foot toes motion, orange: right-foot toes motion). Rows 2-6: time-series of a representative voxel for each task with the PFM-estimated innovation (blue), PFM-estimated activity-inducing (green), and activity-related (i.e., fitted, orange) signals, with their corresponding GLM, PFM, and TA maps on the right (representative voxels indicated with green arrows). Amplitudes are given in signal percentage change (SPC). The maps shown on the right are sampled at the time-points labeled with the red arrows and display the innovation signals at these moments across the whole brain.}
\label{fig:task_maps}
\end{figure}

Figure~\ref{fig:task_maps} depicts the results of the analysis of the Motor
dataset with the PFM and TA algorithms using the BIC selection of $\lambda$ (see
Figure~\ref{fig:task_mad} for results with MAD selection), as well as a
conventional GLM approach. The Activation Time Series (top left), calculated as
the sum of squares of all voxel amplitudes (positive vs. negative) for a given
moment in time, obtained with PFM and TA show nearly identical patterns. These
ATS help to summarize the four dimensional information available in the results
across the spatial domain and identify instances of significant BOLD activity.
The second to sixth rows show the voxel timeseries and the corresponding
activity-related, activity-inducing and innovation signals obtained with PFM
using the BIC criterion of representative voxels in the regions activated in
each of the motor tasks. The TA-estimated time-series are not shown because they
were virtually identical. The maps shown on the right correspond to statistical
parametric maps obtained with the GLM for each motor condition ($p < 0.001$) as
well as the maps of the PFM and TA estimates at the onsets of individual motor
events (indicated with arrows in the timecourses). The estimated
activity-related, activity-inducing and innovation signals clearly reveal the
activity patterns of each condition in the task, as they exhibit a BOLD response
locked to the onset and duration of the conditions. Overall, activity maps of
the innovation signal obtained with PFM and TA highly resemble those obtained
with a GLM for individual events, with small differences arising from the
distinct specificity of the GLM and deconvolution analyses. Notice that the
differences observed with the different approaches to select $\lambda$ based on
the MAD estimate shown in Figure~\ref{fig:rss} are reflected on the ATS shown in
Figure~\ref{fig:task_mad} as well.

\begin{figure}[t!]
    \begin{center}
        \includegraphics[width=\textwidth]{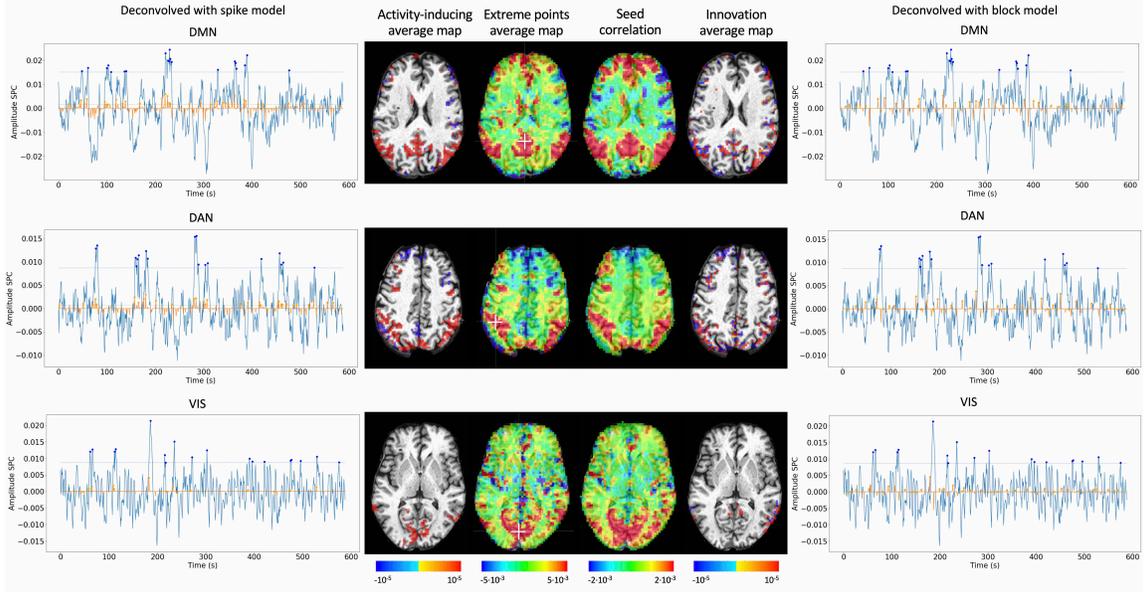}
    \end{center}
    \caption{Average activity-inducing (left) and innovation (right) maps
    obtained from PFM-estimated activity-inducing and innovation signals,
    respectively, using a BIC-based selection of $\lambda$. Time-points selected
    with a 95\textsuperscript{th} percentile threshold (horizontal lines) are
    shown over the average time-series (blue) in the seed region (white cross)
    and the deconvolved signals, i.e., activity inducing (left) and innovation
    (right) signals (orange). Average maps of extreme points and seed
    correlation maps are illustrated in the center.}
\label{fig:caps}
\end{figure}

As an illustration of the insights that deconvolution methods can provide in the
analysis of resting-state data, Figure~\ref{fig:caps} depicts the average
activity-inducing and innovation maps of common resting-state networks obtained
from thresholding and averaging the activity-inducing and innovation signals,
respectively, estimated from the resting-state multiband data using PFM with a
selection of $\lambda$ based on the BIC. The average activity-inducing maps
obtained via deconvolution show spatial patterns of the default mode network
(DMN), dorsal attention network (DAN), and visual network (VIS) that highly
resemble the maps obtained with conventional seed correlation analysis using
Pearson's correlation, and the average maps of extreme points of the signal
(i.e., with no deconvolution). With deconvolution, the average activity-inducing
maps seem to depict more accurate spatial delineation (i.e., less smoothness)
than those obtained from the original data, while maintaining the structure of
the networks. The BIC-informed selection of $\lambda$ yields spatial patterns of
average activity-inducing and innovation maps that are more sparse than those
obtained with a selection of $\lambda$ based on the MAD estimate (see
Figure~\ref{fig:caps_mad}). Furthermore, the spatial patterns of the average
innovation maps based on the innovation signals using the block model yield
complementary information to those obtained with the activity-inducing signal
since iCAPs allow to reveal regions with synchronous innovations, i.e., with the
same upregulating and downregulating events. For instance, it is interesting to
observe that the structure of the visual network nearly disappears in its
corresponding average innovation maps, suggesting the existence of different
temporal neuronal patterns across voxels in the primary and secondary visual
cortices.

%% file: sections/discussion.tex

\section{Discussion and conclusion}

Hemodynamic deconvolution can be formulated using a synthesis- and
analysis-based approach as proposed by PFM and TA, respectively. This work
demonstrates that the theoretical equivalence of both approaches is confirmed in
practice given virtually identical results when the same HRF model and
equivalent regularization parameters are employed. Hence, we argue that
previously observed differences in performance can be explained by specific
settings, such as the HRF model and selection of the regularization parameter
(as shown in Figures~\ref{fig:rss},\ref{fig:rss_comparison} and
\ref{fig:motor_regpaths}), convergence thresholds, as well as the addition of a
spatial regularization term in the spatiotemporal TA formulation
\citep{Karahanoglu2013TotalactivationfMRI}. For instance, the use of PFM with
the spike model in \citep{Tan_2017} was seen not to be ideal due to the
prolonged trials in the paradigm, which better fit the block model as described
here \eqref{eq:pfm_block}. However, given the equivalence of the temporal
deconvolution, incorporating extra spatial or temporal regularization terms in
the optimization problem should not modify this equivalence providing convex
operators are employed. For a convex optimization problem, with a unique global
solution, iterative shrinkage thresholding procedures alternating between the
different regularization terms guarantee convergence; e.g., the generalized
forward-backward splitting \citep{Raguet2013GeneralizedForwardBackward}
algorithm originally employed for TA. Our findings are also in line with the
equivalence of analysis and synthesis methods in under-determined cases (\(N
\leq V\)) demonstrated in \citep{Elad2007Analysisversussynthesis} and
\citep{ortelli2019synthesis}. Still, we have shown that a slight difference in
the selection of the regularization parameter can lead to small differences in
the estimated signals when employing the block model with the MAD selection of
$\lambda$. However, since their regularization paths are equivalent, the
algorithms can easily be forced to converge to the same selection of $\lambda$,
thus resulting in identical estimated signals.

Nevertheless, the different formulations of analysis and synthesis deconvolution
models bring along different kinds of flexibility. One notable advantage of PFM
is that it can readily incorporate any HRF as part of the synthesis operator
\citep{Elad2007Analysisversussynthesis}, only requiring the sampled HRF at the
desired temporal resolution, which is typically equal to the TR of the
acquisition. Conversely, TA relies upon the specification of the discrete
differential operator that inverts the HRF, which needs to be derived either by
the inverse solution of the sampled HRF impulse response, or by discretizing a
continuous-domain differential operator motivated by a biophysical model. The
more versatile structure of PFM allows for instance an elegant extension of the
algorithm to multi-echo fMRI data
\citep{CaballeroGaudes2019deconvolutionalgorithmmulti} where multiple
measurements relate to a common underlying signal. Therefore, the one-to-many
synthesis scenario (i.e., from activity-inducing to several activity-related
signals) is more cumbersome to express using TA; i.e., a set of differential
operators should be defined and the differences between their outputs
constrained. Conversely, the one-to-many analysis scenario (i.e., from the
measurements to several regularizing signals) is more convenient to be expressed
by TA; e.g., combining spike and block regularizers. While the specification of
the differential operator in TA only indirectly controls the HRF, the use of the
derivative operator to enforce the block model, instead of the integrator in
PFM, impacts positively the stability and rate of the convergence of the
optimization algorithms. Moreover, analysis formulations can be more suitable
for online applications that are still to be explored in fMRI data, but are
employed for calcium imaging deconvolution \citep{Friedrich_2017,Jewell_2019},
and which have been applied for offline calcium deconvolution
\citep{Farouj2020DeconvolutionSustainedNeural}.

Deconvolution techniques can be used before more downstream analysis of brain
activity in terms of functional network organization as they estimate
interactions between voxels or brain regions that occur at the activity-inducing
level, and are thus less affected by the slowness of the hemodynamic response
compared to when the BOLD signals are analyzed directly. In addition,
deconvolution approaches hold a close parallelism to recent methodologies aiming
to understand the dynamics of neuronal activations and interactions at short
temporal resolution and that focus on extreme events of the fMRI signal
\citep{Lindquist_2007}. As an illustration, Figure 6 shows that the innovation-
or activity-inducing CAPs computed from deconvolved events in a single
resting-state fMRI dataset closely resemble the conventional CAPs computed
directly from extreme events of the fMRI signal
\citep{Liu2013Timevaryingfunctional,Liu2013Decompositionspontaneousbrain,
Liu2018Coactivationpatterns,cifre2020revisiting,Cifre2020Furtherresultswhy,
Zhang2020relationshipBOLDneural,Tagliazucchi2011,Tagliazucchi2012,
Tagliazucchi2016,Rolls2021}. Similarly, we hypothesize that these extreme events
will also show a close resemblance to intrinsic ignition events
\citep{Deco2017a,Deco2017}. As shown in the maps, deconvolution approaches can
offer a more straightforward interpretability of the activation events and
resulting functional connectivity patterns. Here, CAPs were computed as the
average of spatial maps corresponding to the events of a single dataset. Beyond
simple averaging, clustering algorithms (e.g., K-means and consensus clustering)
can be employed to discern multiple CAPs or iCAPs at the whole-brain level for a
large number of subjects. Previous findings based on iCAPs have for instance
revealed organizational principles of brain function during rest
\citep{Karahanoglu2015Transientbrainactivity} and sleep \citep{tarun2101} in
healthy controls, next to alterations in 22q11ds \citep{zoller1902} and multiple
sclerosis \citep{bommarito2101p}. Next to CAPs-inspired approaches, dynamic
functional connectivity has recently been investigated with the use of
co-fluctuations and edge-centric techniques
\citep{Faskowitz2020,Esfahlani2020Highamplitudecofluctuations,Jo2021,Sporns2021,Oort2018}.
The activation time series shown in Figure 5 aim to provide equivalent
information to the root of sum of squares timecourses used in edge-centric
approaches, where timecourses with peaks delineate instances of significant
brain activity. Future work could address which type of information is redundant
or distinct across these frameworks. In summary, these examples illustrate that
deconvolution techniques can be employed prior to other computational approaches
and could serve as an effective way of denoising the fMRI data. We foresee an
increase in the number of studies that take advantage of the potential benefits
of using deconvolution methods prior to functional connectivity analyses.

In sum, hemodynamic deconvolution approaches using sparsity-driven
regularization are valuable tools to complete the fMRI processing pipeline.
Although the two approaches examined in detail here provide alternative
representations of the BOLD signals in terms of innovation and activity-inducing
signals, their current implementations have certain limitations, calling for
further developments or more elaborate models, where some of them have been
initially addressed in the literature. One relevant focus is to account for the
variability in HRF that can be observed in different regions of the brain.
First, variability in the temporal characteristics of the HRF can arise from
differences in stimulus intensity and patterns, as well as with short
inter-event intervals like in fast cognitive processes or experimental designs
\citep{Yesilyurt2008DynamicsnonlinearitiesBOLD,
Sadaghiani2009Neuralactivityinduced,Chen2021Investigatingmechanismsfast,Polimeni2021Imagingfasterneural}.
Similarly, the HRF shape at rest might differ from the canonical HRF commonly
used for task-based fMRI data analysis. A wide variety of HRF patterns could be
elicited across the whole brain and possible detected with sufficiently large
signal-to-noise ratio, e.g., \citep{GonzalezCastillo2012Wholebraintime} showed
two gamma-shaped responses at the onset and the end of the evoked trial,
respectively. This unique HRF shape would be deconvolved as two separate events
with the conventional deconvolution techniques. The impact of HRF variability
could be reduced using structured regularization terms along with multiple basis
functions \citep{Gaudes2012Structuredsparsedeconvolution} or procedures that
estimate the HRF shape in an adaptive fashion in both analysis
\citep{Farouj2019BoldSignalDeconvolution} and synthesis formulations
\citep{cherkaoui:hal-03005584}.

Another avenue of research consists in leveraging spatial information by
adopting multivariate deconvolution approaches that operate at the whole-brain
level, instead of working voxelwise and beyond regional regularization terms
(e.g. as proposed in \citealt{Karahanoglu2013TotalactivationfMRI}). Operating at
the whole-brain level would open the way for methods that consider shared
neuronal activity using mixed norm regularization terms
\citep{urunuela-tremino_2019} or can capture long-range neuronal cofluctuations
using low rank decompositions \citep{cherkaoui:hal-03005584}. For example,
multivariate deconvolution approaches could yield better localized activity
patterns while reducing the effect of global fluctuations such as respiratory
artifacts, which cannot be modelled at the voxel level \citep{Urunuela_2021}.

Similar to solving other inverse problems by means of regularized estimators,
the selection of the regularization parameter is critical to correctly estimate
the neuronal-related signal. Hence, methods that take advantage of a more robust
selection of the regularization parameter could considerably yield more reliable
estimates of the neuronal-related signal. For instance, the stability selection
\citep{Meinshausen2010Stabilityselection,Urunuela2020StabilityBasedSparse}
procedure could be included to the deconvolution problem to ensure that the
estimated coefficients are obtained with high probability. Furthermore, an
important issue of regularized estimation is that the estimates are biased with
respect to the true value. In that sense, the use of non-convex
\(\ell_{p,q}\)-norm regularization terms (e.g., \(p < 1\)) can reduce this bias
while maintaining the sparsity constraint, at the cost of potentially converging
to a local minima of the regularized estimation problem. In practice, these
approaches could avoid the optional debiasing step that overcomes the shrinkage
of the estimates and obtain a more accurate and less biased fit of the fMRI
signal
\citep{Gaudes2013Paradigmfreemapping,CaballeroGaudes2019deconvolutionalgorithmmulti}.
Finally, cutting-edge developments on physics-informed deep learning techniques
for inverse problems \citep{Akcakaya2021,Monga2021,Ongie2020,Cherkaoui_2020}
could be transferred for deconvolution by considering the biophysical model of
the hemodynamic system and could potentially offer algorithms with reduced
computational time and more flexibility.

%% file: sections/supplementary.tex
\pagebreak
\begin{center}
\textbf{\large Supplementary Material for Hemodynamic Deconvolution Demystified: Sparsity-Driven Regularization at Work}
\end{center}
\setcounter{equation}{0}
\setcounter{figure}{0}
\setcounter{table}{0}
\setcounter{page}{1}
\makeatletter
\renewcommand{\theequation}{S\arabic{equation}}
\renewcommand{\thefigure}{S\arabic{figure}}
\renewcommand{\bibnumfmt}[1]{[S#1]}
\renewcommand{\citenumfont}[1]{S#1}

\begin{figure*}[h!]
    \begin{center}
        \includegraphics[width=\textwidth]{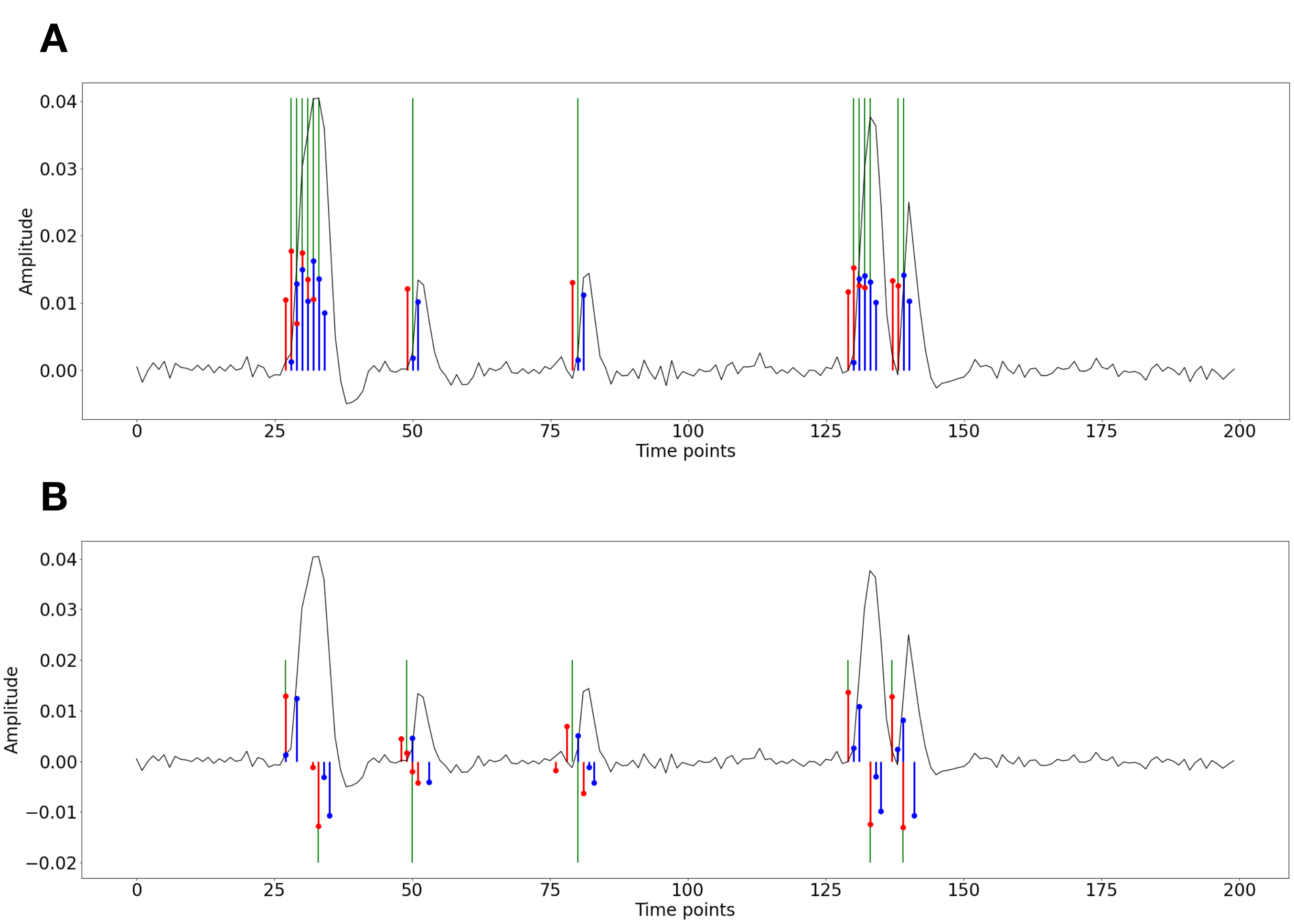}
    \end{center}
    \caption{Activity-inducing (A) and innovation (B) signals estimated with
    PFM (red) and TA (blue) using their built-in HRF as opposed to using the
    same. The black line depicts the simulated signal, while the green lines
    indicate the onsets of the simulated neuronal events. X axis shows time
    in TRs.}
\label{fig:hrf_differences}
\end{figure*}

\begin{figure*}[h!]
    \begin{center}
        \includegraphics[width=\textwidth]{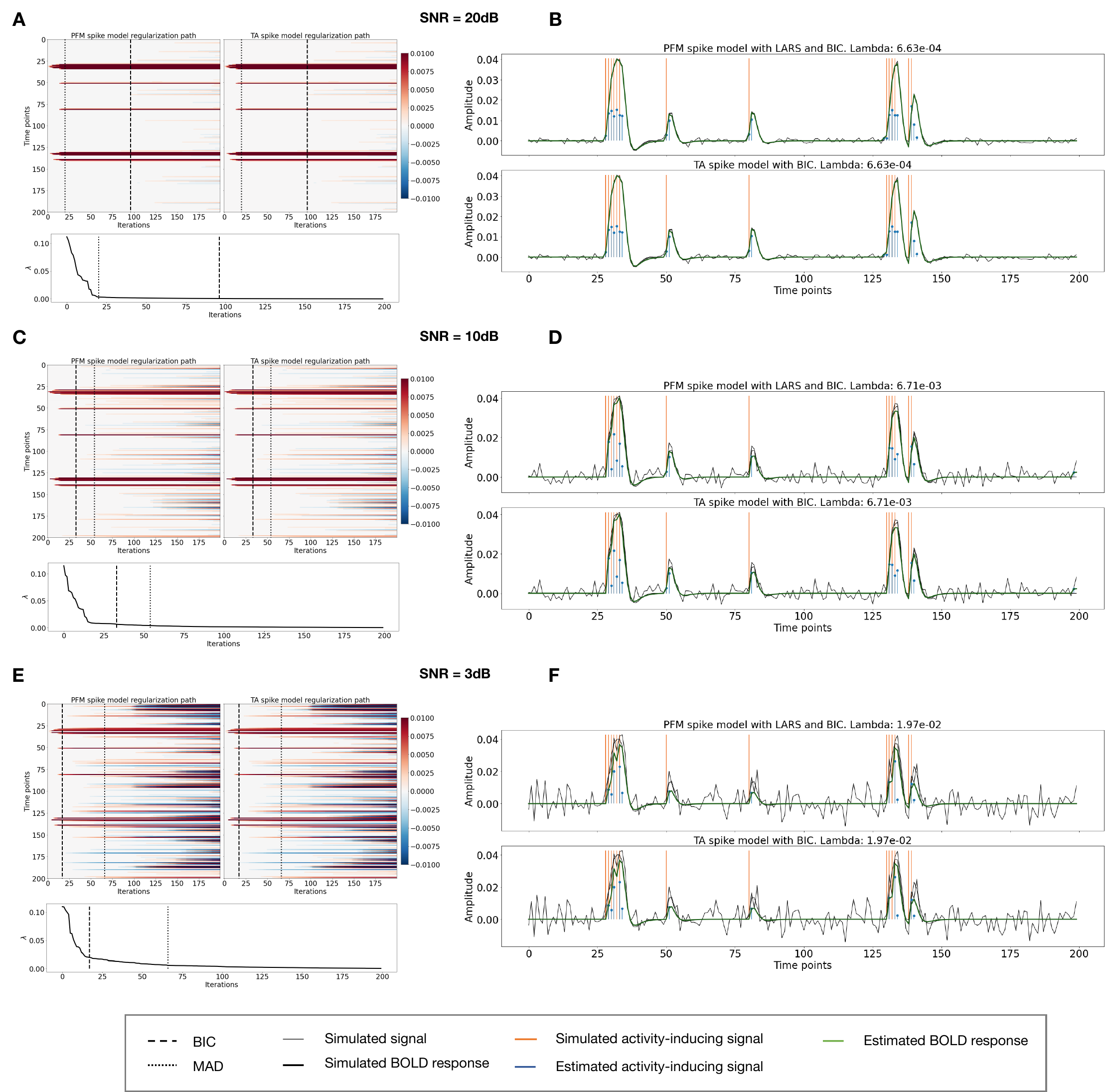}
    \end{center}
    \caption{Spike model simulations. (Left) Heatmap of the regularization
    paths of the activity-inducing signal estimated with PFM and TA as a
    function of $\lambda$ (increasing number of iterations in x-axis),
    whereas each row in the y-axis shows one time-point. Vertical lines
    denote iterations corresponding to the Akaike and Bayesian Information
    Criteria (AIC and BIC) optima. (Right) Estimated activity-inducing (blue)
    and activity-related (green) signals when set based on BIC. All estimates
    are identical, regardless of SNR.}
\label{fig:path_spike}
\end{figure*}

\begin{figure*}[h!]
    \begin{center}
        \includegraphics[width=\textwidth]{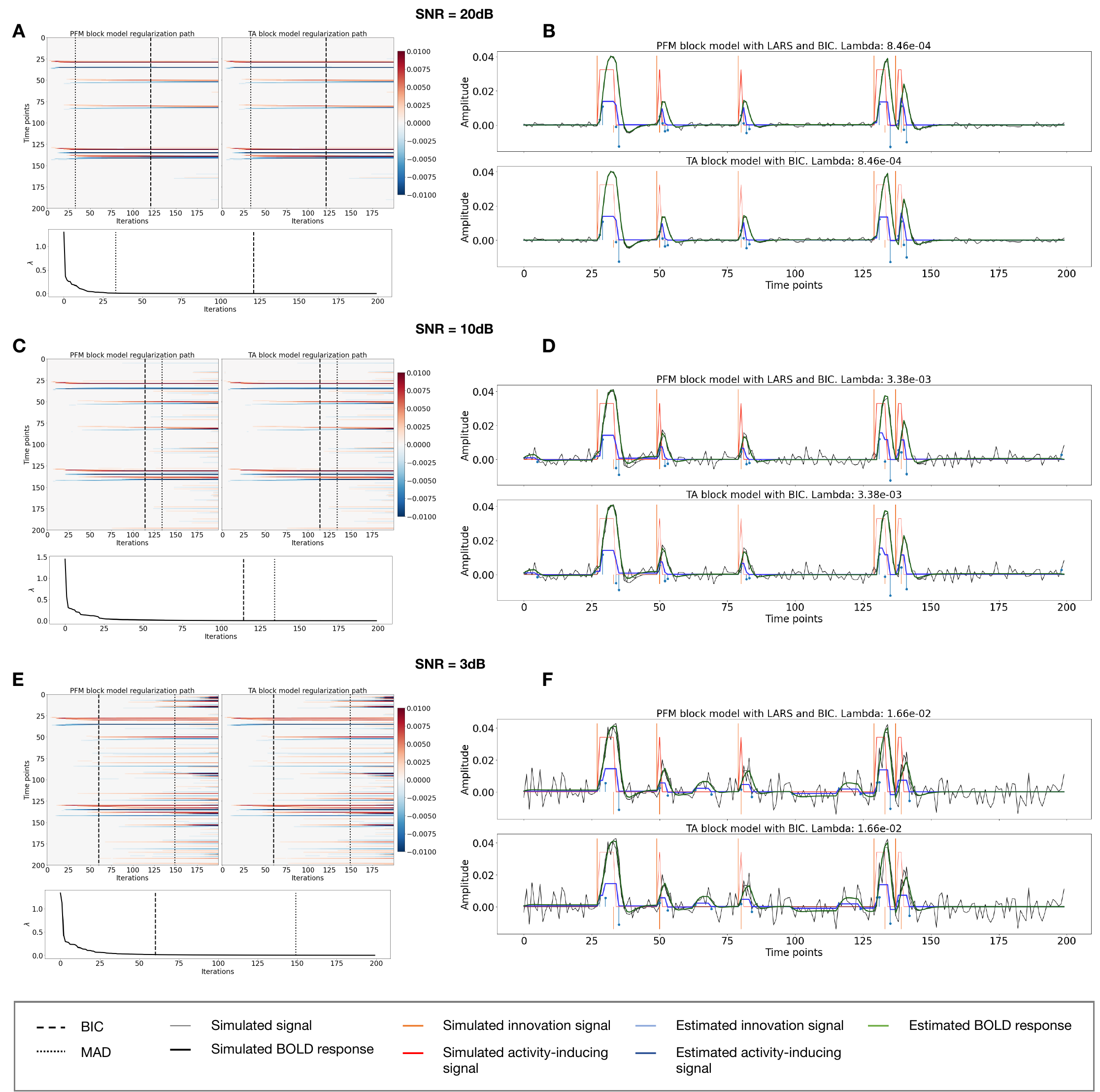}
    \end{center}
    \caption{Block model simulations. (Left) Heatmap of the regularization
    paths of the innovation signal estimated with PFM and TA as a function
    of $\lambda$ (increasing number of iterations in x-axis), whereas
    each row in the y-axis illustrates one time-point. Vertical lines denote
    iterations corresponding to the Akaike and Bayesian Information Criteria
    (AIC and BIC) optima. (Right) Estimated innovation (blue) and
    activity-related (green) signals when is set based on BIC. All the
    estimates are identical when compared between the PFM and TA cases,
    regardless of SNR.}
\label{fig:path_block}
\end{figure*}


\begin{figure*}[h!]
    \begin{center}
        \includegraphics[width=\textwidth]{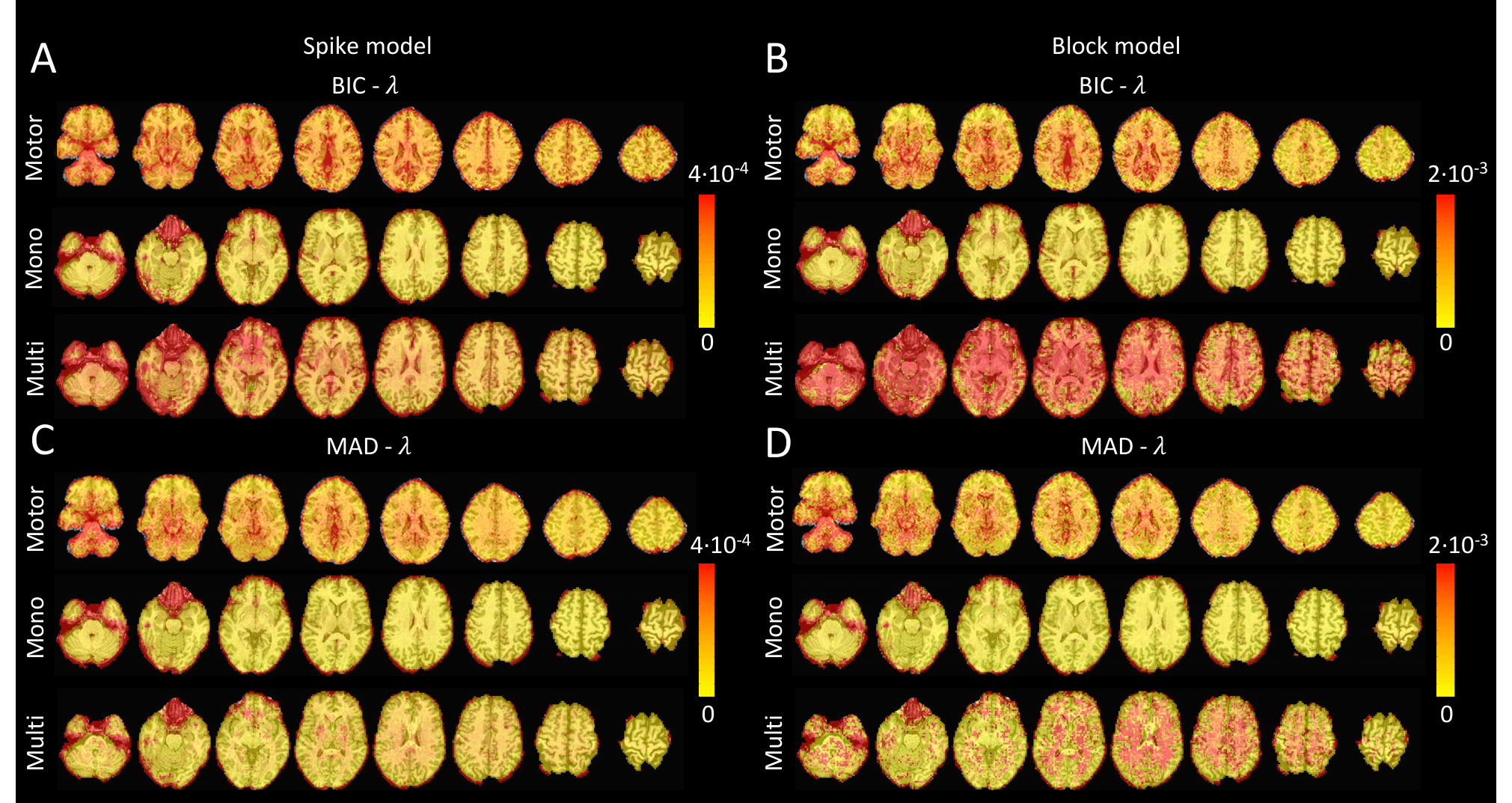}
    \end{center}
    \caption{Values of $\lambda$ across the different voxels in the brain
    used to estimate (A) the activity-inducing signal (spike model) and (B)
    the innovation signal (block model) with the BIC selection, as well as
    (C) the activity-inducing signal (block model) and (D) the innovation
    signal (block model) with a MAD-based selection. The $\lambda$ maps are
    shown for the three experimental fMRI datasets: the motor task (Motor),
    the monoband resting-state (Mono), and the multiband resting-state (Multi)
    datasets.}
\label{fig:lambdas}
\end{figure*}

\begin{figure*}[h!]
    \begin{center}
        \includegraphics[width=\textwidth]{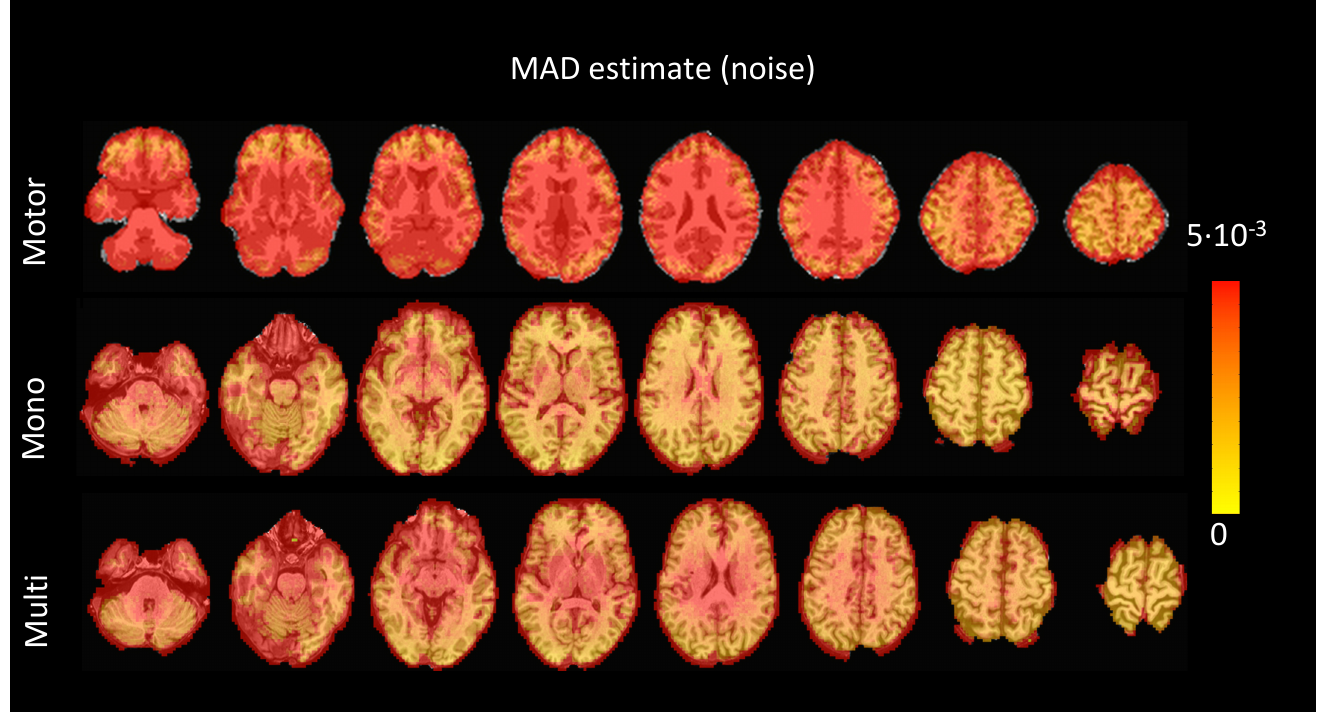}
    \end{center}
    \caption{Values of the MAD estimate of standard deviation of the noise
    across the different voxels in the brain for the three experimental fMRI
    datasets: the motor task (Motor), the monoband resting-state (Mono), and
    the multiband resting-state (Multi) datasets.}
\label{fig:mad_estimate}
\end{figure*}

\begin{figure*}[h!]
    \begin{center}
        \includegraphics[width=\textwidth]{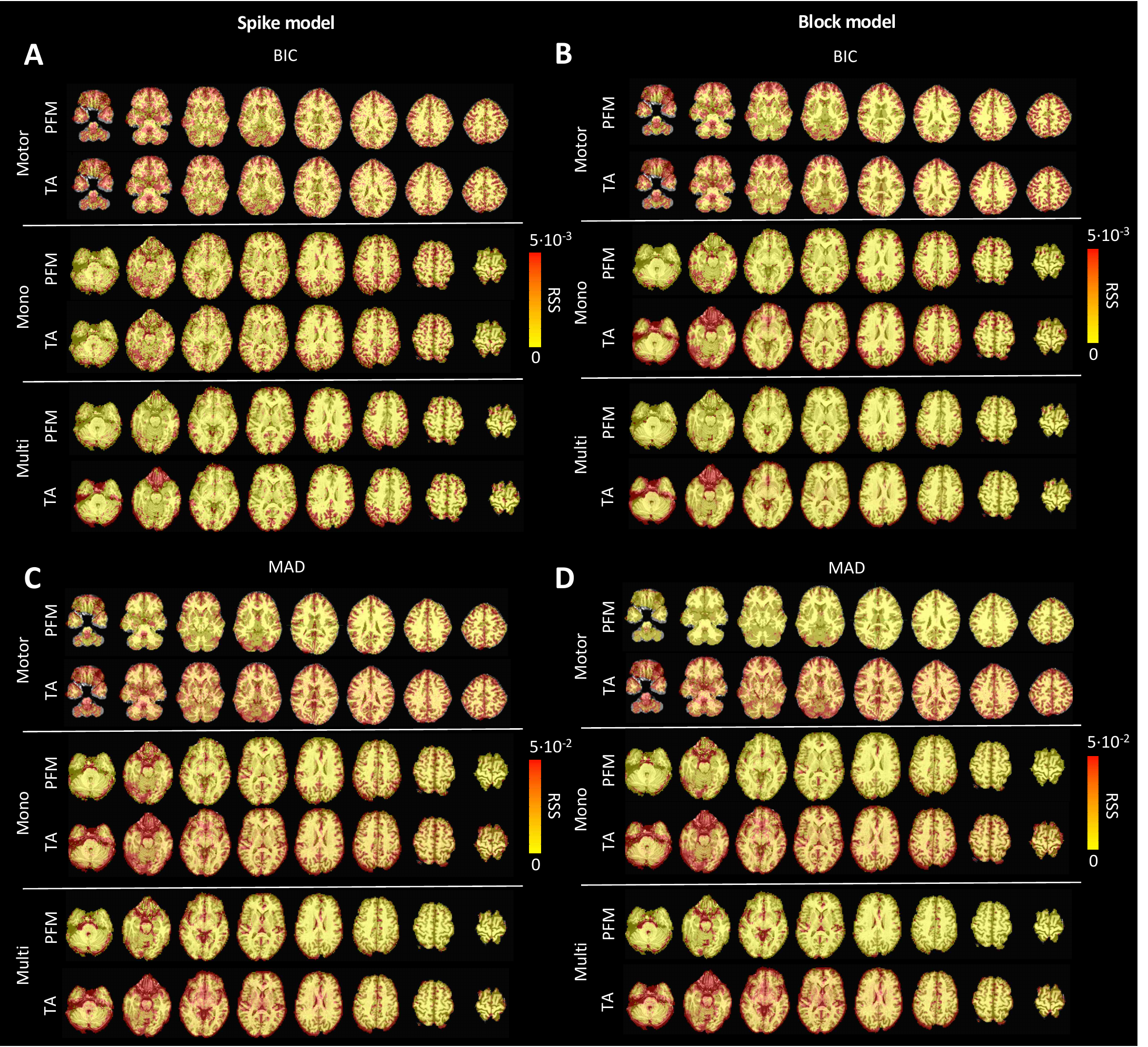}
    \end{center}
    \caption{Root sum of squares (RSS) comparison between Paradigm Free Mapping
    and Total Activation for the three experimental fMRI datasets: the motor
    task (Motor), the monoband resting-state (Mono), and the multiband
    resting-state (Multi) datasets. RSS maps are shown for the spike (left) and
    block (right) models solved with a selection of $\lambda$ based on the BIC
    (top) and MAD (bottom) criteria.}
\label{fig:rss_comparison}
\end{figure*}

\begin{figure*}[h!]
    \begin{center}
        \includegraphics[width=\textwidth]{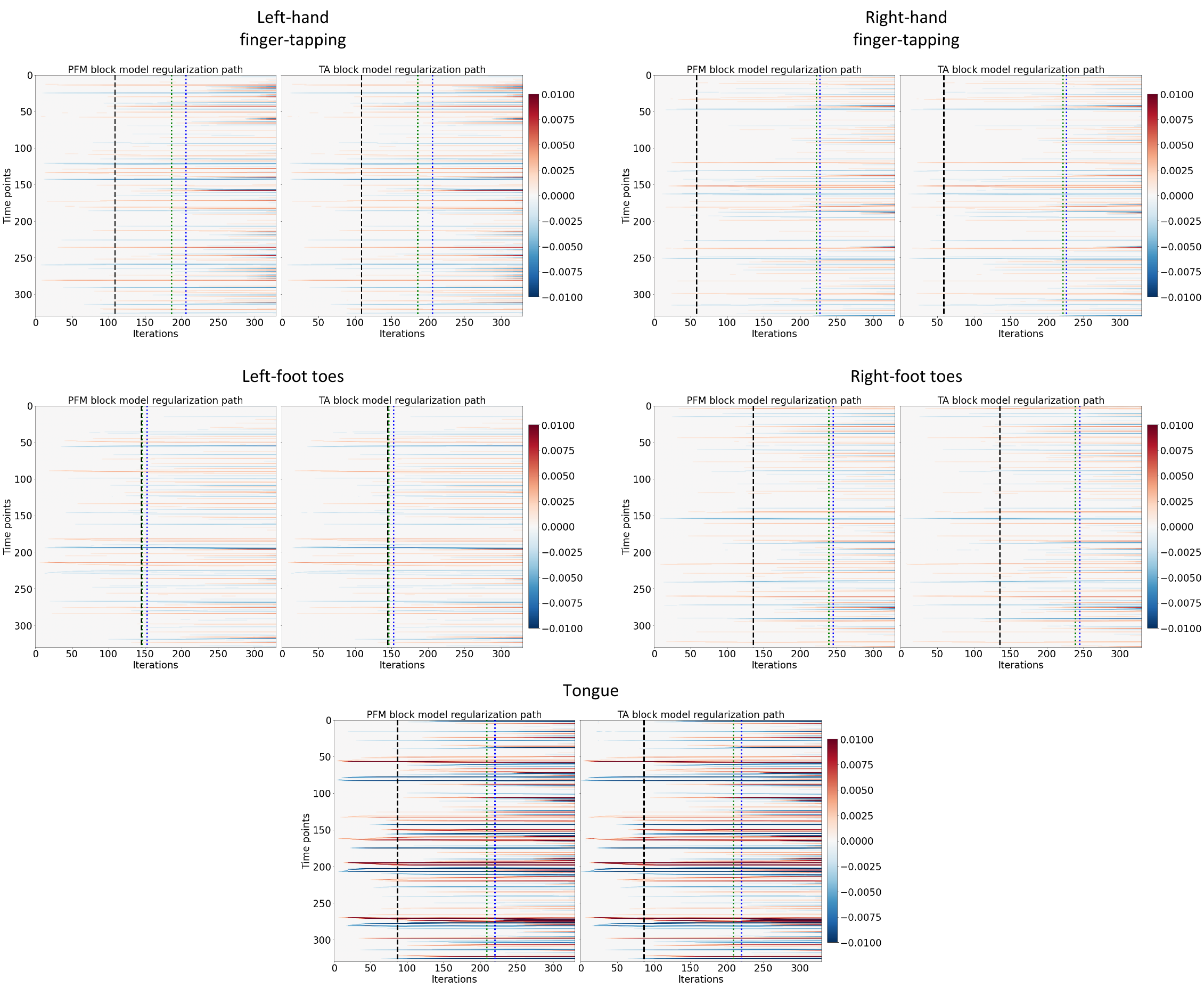}
    \end{center}
    \caption{Regularization paths of the innovation signal estimated with
    PFM and TA as a function of $\lambda$ (increasing number of iterations in
    x-axis, whereas each row in the y-axis shows one time-point) for the
    representative voxels of the motor task shown in Figure \ref{fig:task_maps}.
    Vertical lines denote selections of $\lambda$ corresponding to the BIC
    (black), MAD based on LARS residuals (blue) and MAD based on FISTA residuals
    (green) optima.}
\label{fig:motor_regpaths}
\end{figure*}

\begin{figure*}[h!]
    \begin{center}
        \includegraphics[width=\textwidth]{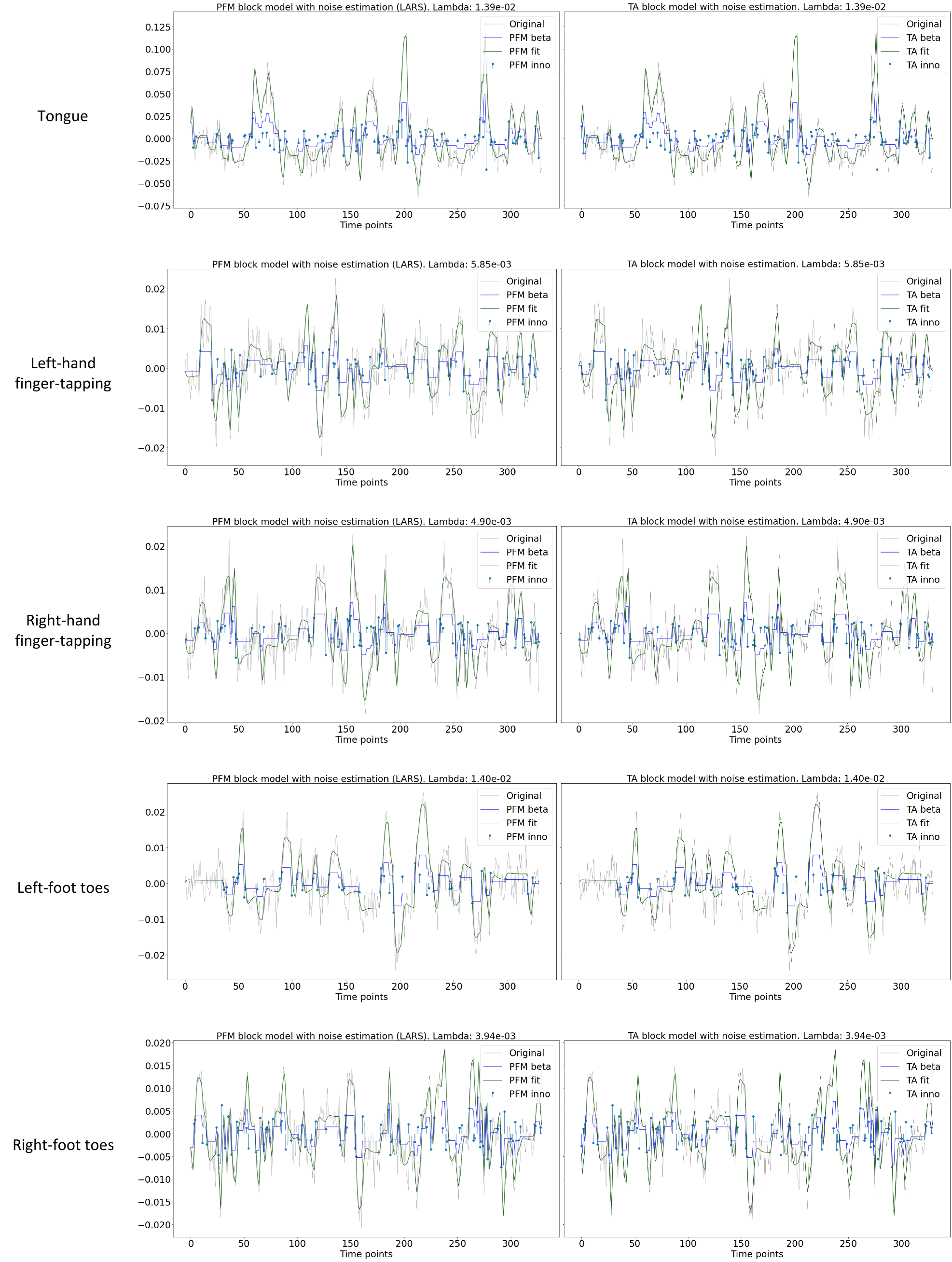}
    \end{center}
    \caption{Estimated innovation signal (blue) and activity-related signal
    (green) for the representative voxels of the motor task shown in
    Figure~\ref{fig:task_maps} with the MAD selection of $\lambda$ made by TA,
    i.e., employing the same $\lambda$ with both PFM and TA.}
\label{fig:mad_inno_ts}
\end{figure*}

\begin{figure*}[h!]
    \begin{center}
        \includegraphics[width=\textwidth]{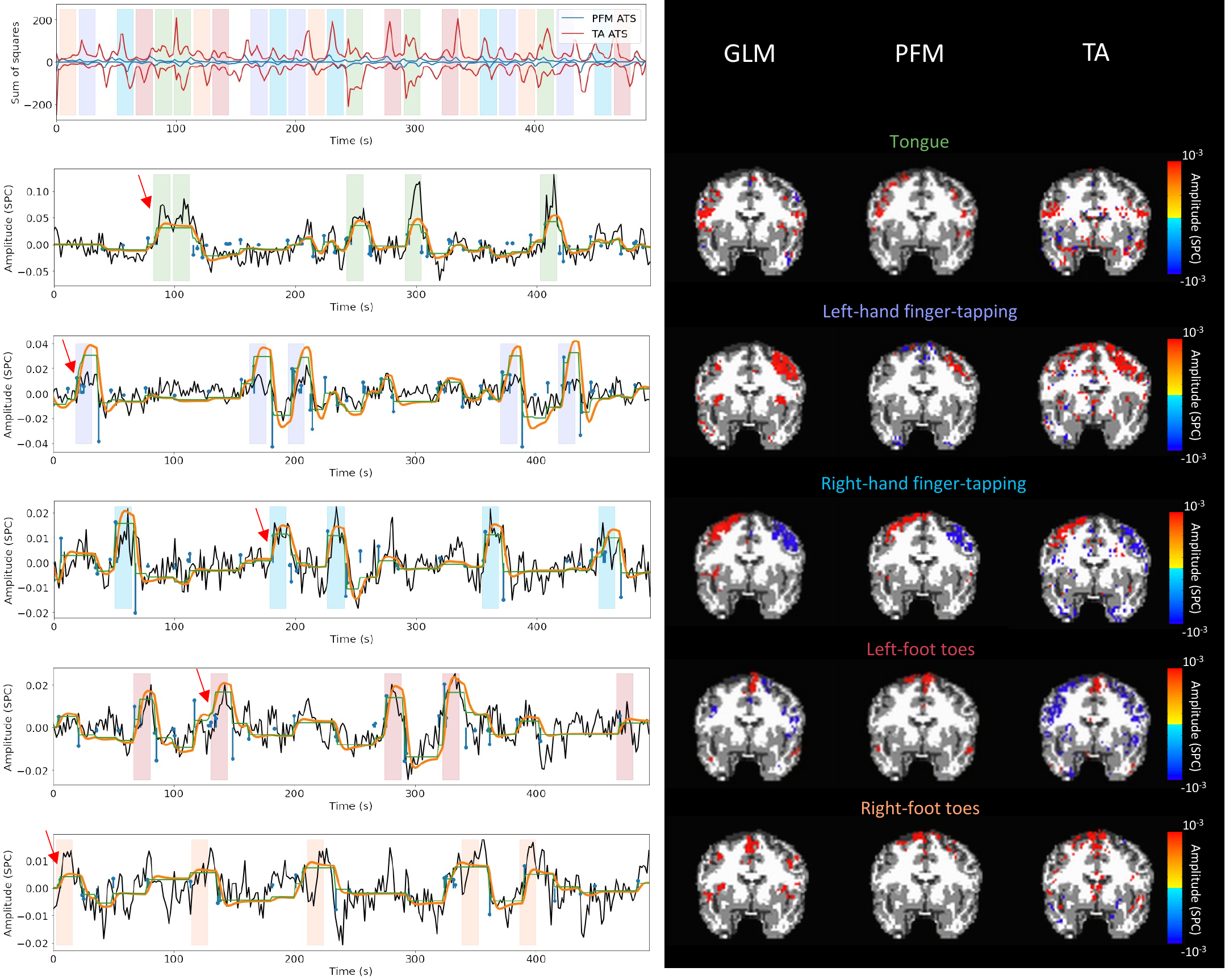}
    \end{center}
    \caption{Activity maps of the motor task using a seletion of $\lambda$ based
    on the MAD estimate. Row 1: Activation time-series of the innovation signals
    estimated by PFM (in blue) or TA (in red) calculated as the sum of squares
    of all voxels at every timepoint. Positive-valued and negative-valued
    contributions were separated into two distinct time-courses. Color-bands
    indicate the onset and duration of each condition in the task (green:
    tongue, purple: left-hand finger-tapping, blue: right-hand finger-tapping,
    red: left-foot toes, orange: right-foot toes). Rows 2-6: time-series of a
    representative voxel for each task with the PFM-estimated innovation (blue),
    PFM-estimated activity-inducing (green), and activity-related (i.e., fitted,
    orange) signals, with their corresponding GLM, PFM, and TA maps on the
    right. The maps shown on the right are sampled at the time-point labeled
    with the red arrows and display the innovation signals at that moment across
    the whole brain.}
\label{fig:task_mad}
\end{figure*}

\begin{figure*}[h!]
    \begin{center}
        \includegraphics[width=\textwidth]{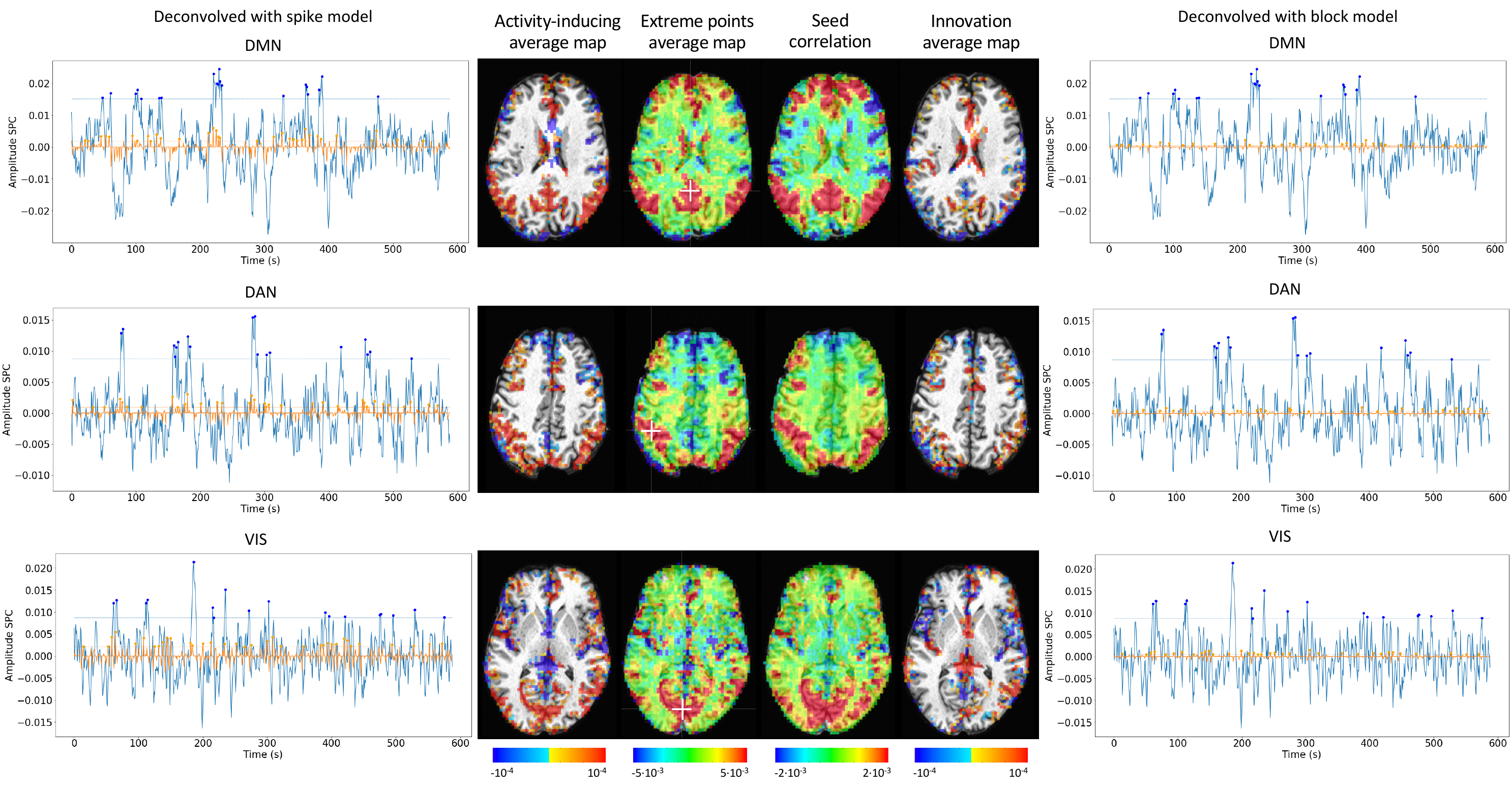}
    \end{center}
    \caption{Activity-inducing CAPs (left) and innovation CAPs (right) obtained
    with the PFM-estimated activity-inducing and innovation signals
    respectively, using a MAD-based selection of $\lambda$. Time-points selected
    with a 95th percentile threshold are shown over the average time-series
    (blue) in the seed region (white-cross) and the deconvolved signal (orange).
    CAPs and seed correlation maps are illustrated in the center.}
\label{fig:caps_mad}
\end{figure*}